# Asymmetric Bidirectional Quantum Teleportation: Arbitrary bi-modal Information State

Ankita Pathak[*1], Madan Singh Chauhan and Ravi S. Singh[*2]
Photonic Quantum-Information and Quantum Optics Group, Department of Physics,
Deen Dayal Upadhyaya Gorakhpur University, Gorakhpur (U.P.), 273009, India.

**e-mail:** [1]pathak18.phy@gmail.com; [2]yesora27@gmail.com

## Abstract

Optical coherent states are experimentally realizable continuous variable quantum states of which preparation by lasers, as well as its manipulation and monitoring by linear optical gadgets are well established. We propose a strategy to send an arbitrary superposition of four-component bi-modal entangled coherent states from a sender to a receiver who, simultaneously, tries to transmit an unknown Schrodinger Cat coherent-state to sender via employing a cluster consisting of 'three' superposition of two-component bi-modal entangled coherent states as the quantum channel and utilizing linear optical gadgets. Heralded detections of photons in laboratories of sender and receiver followed by classical communications of even/odd number of photons and local unitary operations, impeccably, accomplishes simultaneous faithful asymmetric bidirectional quantum teleportation with one-eighth of probability of success. It is seen that not all detection events implement the protocol and, therefore, one has to locally apply displacement operator, a necessary evil. We analyze near-faithful partial asymmetric bidirectional quantum teleportation and associated probability of success therein. We demonstrated that, for an intense coherent optical field, fidelities approach unity.

**Keywords**: Bidirectional Quantum Teleportation, Schrodinger Cat states, Entangled Coherent States, Linear Optics.

## 1. Introduction

Quantum Teleportation (QT) is a fundamental protocol in quantum communication discovered by Bennet et al. [1] in discrete-variable regime. Continuous-variable version of QT has been demonstrated by Vaidman [2] via utilizing the conception of nonlocal measurements [3,4], which, later on, developed by Braunstein and Kimble [5] through Wigner distribution function route. Quantum communication has witnessed myriad variants such as quantum cryptography [6,7], super dense coding [8], quantum secure direct communication [9], quantum secret sharing [10], remote state preparation [11,13], port-based QT [14,15], counterfactual quantum communication [16], Catalytic QT [17-19]. QT is experimentally realized in various formidable physical systems such as photonic states [20-25], optical quantum modes [26-28], nuclear magnetic resonance [29], atomic ensembles [30-32], trapped atoms [33-35] or solid-state systems [36-38]. Moreover, QT is being unified with novel prevalent concepts to bring forth novel schemes such as quantum telecomputing [39-41], quantum telecloning [42-44], quantum telefilters and telemirrors [45], quantum broadcasting [46,47], quantum tele-amplification [48], quantum digital signature [49], to name a few.

Cochrane et. al. [50] have utilized, for the first time, even and odd coherent states [51] to encode logical qubits. Ralph et.al. [52] have advocated that coherent states, itself, may be mapped as logical-qubits by $|0\rangle_L \rightarrow|\alpha\rangle$ and $|1\rangle_L\rightarrow|-\alpha\rangle$, α is, here, taken real, where $|\alpha\rangle$ and $|-\alpha\rangle$ are π-phase apart optical coherent states. Obviously, although $|\alpha\rangle$ and $|-\alpha\rangle$ are linearly independent non-orthogonal vectors embedded individually in unbounded Hilbert space but their superposition, namely, Schrodinger Cat states, are in two-dimensional Hilbert space.

van Enk and Hirota [53] have designed a scheme by using only linear optical elements such as beam splitters, phase shifters and photon number resolving detectors, in which Schrodinger Cat coherent-states is teleported to non-locally stationed Bob via bi-modal maximally Bell-coherent states [54] as quantum channel. Ba An [55] have worked out a scheme for teleporting an arbitrary superposed coherent state within a network. Wang [56] have generalized the van Enk and Hirota's protocol by in which an unknown Bell coherent-state is transmitted by GHZ-coherent states. Furthermore, teleportation of an arbitrary bi-modal four component entangled coherent states have been proposed by Liao and Kuang [54] and Phien and Ba An [57], Prakash et.al. [58].

Karlsson and Bourennane [59] generalized standard QT protocol [1] by inducting a third party, Charlie playing the role of controller in the scheme, and, therefore, notion of controlled quantum teleportation appears. Unidirectional controlled QT witnessed many variants by inducting 'multi-parties', multi qubits as information encoding and demanding security-issues by developing novel communication schemes in discrete variable regime [60-63]. Notably, controlled QT employing GHZ-coherent-states as quantum channel is brought out by Pandey et.al. [64] for transmitting unknown superposed coherent states from sender Alice to receiver Bob controlled by Charlie. Bidirectional QT protocol has appeared in work of Huelga et.al. [65], who have, while investigating implementation of arbitrary unitary operation upon distantly located quantum system, proven that the required resources for the same are those of implementing bidirectional QT. Furthermore, bidirectional controlled quantum teleportation is proposed in DV-regime by Zha et.al. [66], of which generalization is proposed by Shukla et.al. [67]. Bidirectional QT has received a renewed interests through investigating optimal input states and quantifying performance [68,69]. Bidirectional controlled quantum teleportation is a three-party scheme employing shared multi-particle maximally or mixed entangled states as quantum channel in which communication of quantum-information encoded either in single qubit or in entangled qubits possessed by sender (receiver) are exchanged with the assistance of third party acting as 'controller'. A sudden increase of interests is witnessed by many proposals on bidirectional QT [70-72] and its ramification in discrete variable regime on the basis of 'to be teleported single (multi) qubit information-state' at sender (receiver) end and, also various kinds of multi-qubit entangled state utilized as quantum channel [73,74]. Communication in quantum network, indispensably, necessitates exchange of quantum states possessed by parties such as asymmetric bidirectional quantum teleportation [75-77], cyclic quantum teleportation [78,79], cyclic 'controlled' quantum teleportation [80-82], multidirectional quantum teleportation in discrete variable regime [83,84].

Recently, applying ideal linear optical devices, a symmetric bidirectional cyclic quantum teleportation in which information-states at sender (receiver) station are encoded arbitrary superposed optical coherent states and penta-modal cluster-type entangled coherent states is employed as quantum channel [85]. Aliloute et.al. [86] designed a critically incorrect symmetric bidirectional QT scheme using tri-modal GHZ-entangled coherent states, but they made erroneous task by mixing Alice and Bob quantum information with a beam splitter destroying the encoding. Quite recently, we introduced a theoretical protocol for asymmetric bidirectional quantum teleportation. In this scheme, a sender transmits bi-modal entangled coherent states to a receiver, who simultaneously sends Schrödinger Cat coherent-states back to the sender [87]. Following the same idea, we have generalized our scheme to teleport an arbitrary four-component bi-modal entangled coherent states, continuous variable superposition information state, from Alice (sender) to Bob(receiver) who, simultaneously, endeavors to transmit Schrodinger Cat coherent-state to sender via employing a cluster Bell coherent-states as the quantum channel and utilizing linear optical gadgets. Heralding technique followed by classical communications of photon-counts and local unitary operations realized the simultaneous faithful and near-faithful asymmetric bidirectional quantum teleportation. One may notice that, for an intense coherent optical field, fidelity of near-faithful asymmetric bidirectional quantum teleportation approaches unit value.

The paper is organized in following Sections. Section (2) describes our scheme of bidirectional quantum teleportation with a continuous variable superposition state in detail. The analysis for probability of success and fidelity of near-faithful/ faithful is provided in Section (3).Finally, conclusions are sketched based upon communication-complexity of proposed scheme and future prospects.

## 2. Bidirectional Quantum Teleportation Protocol

Alice has to teleport an arbitrary superposition of four-component bi-modal entangled coherent state,

$$|\psi\rangle_{AA'} = N_{AA'}(A_0|\alpha,\alpha\rangle + A_1|\alpha,-\alpha\rangle + A_2|-\alpha,\alpha\rangle + A_3|-\alpha,-\alpha\rangle)\,(1)$$

wherenormalization constant,

$N_{AA'} = \left[\sum_{i=0}^{3}|A_i|^2 + 2e^{-2|\alpha|^2}\mathrm{Re}(A_0^*A_2 + A_1^*A_3 + A_0^*A_1 + A_2^*A_3) + 2e^{-4|\alpha|^2}\mathrm{Re}(A_0^*A_3 + A_1^*A_2)\right]^{-1/2}$, $\alpha$ is taken to be real for the sake of simplicity and without loss of generality, Re ≡ Real and, also, Bob attempts to send a mono-modal superposition of coherent states encoded in Schrodinger Cat coherent-state, experimentally prepared [88, 89],

$$|\psi\rangle_B = N_B(B_0|\alpha\rangle + B_1|-\alpha\rangle)\,(2)$$

where, $N_B$is a normalization constant $N_B = \left[\sum_{i=0}^{1}|B_i|^2 + 2e^{-2|\alpha|^2}\mathrm{Re}(B_1^*B_0)\right]^{-1/2}$ ,Since Alice's 'to be transmitted' state is an arbitrary bi-modal coherent state and that from Bob to Alice is mono-modal Schrodinger coherent Cat-states, the teleportation may, evidently, be termed as Asymmetric Bidirectional Quantum Teleportation (ABQT). To realize ABQT protocol, we employed hexa-modal entangled coherent states as a quantum channel,

$$|\psi\rangle_{1,2,3,4,5,6} = N_C|\Omega\rangle_{1,4}\otimes|\Omega\rangle_{2,5}\otimes|\Omega\rangle_{3,6}, \qquad (3)$$

with normalization constant, $N_c = [8(1 + e^{-12|\alpha|^2} + 3e^{-8|\alpha|^2} + e^{-6|\alpha|^2} + 2e^{-4|\alpha|^2}]^{-1/2}$ .To prepare quantum channel, Eq.(3), it suffices to prepare Bell coherent-states, say,

$$|\Omega\rangle_{i,j} = \hat{B}\left(\tilde{N}|\alpha\sqrt{2}\rangle + |-\alpha\sqrt{2}\rangle\right)_x \otimes |0\rangle_y = \tilde{N}(|\alpha,\alpha\rangle + |-\alpha,-\alpha\rangle)_{i,j} \xrightarrow{\hat{P}_j(\pi)=e^{(-i\pi\hat{a}_j^\dagger\hat{a}_j)}} = \tilde{N}(|\alpha,-\alpha\rangle + |-\alpha,\alpha\rangle)_{i,j},$$

where $\hat{B}$ corresponds to "symmetric beam splitter with phase shifter" (BPS) operation, such that

$$\hat{B}\left(|\beta\rangle_x \otimes |\gamma\rangle_y\right) = \left|\frac{\beta+\gamma}{\sqrt{2}}\right\rangle_i, \left|\frac{\beta-\gamma}{\sqrt{2}}\right\rangle_j \qquad (4)$$

The protocol is strategized by distributing modes of quantum channel; Eq. (3) such that Alice preserves modes 1,2 and 6 while Bob possesses modes 3, 4 and 5, respectively. The global state can, be expressed as, $|\varphi\rangle_{AA'B,1,2,3,4,5,6} = N_T\left(|\psi\rangle_{AA'} \otimes |\psi\rangle_B \otimes |\psi\rangle_{123456}\right)$

$$|\varphi\rangle_{A,A',B,1,2,3,4,5,6} = N_T(A_0B_0|\alpha,\alpha,\alpha\rangle + A_0B_1|\alpha,\alpha,-\alpha\rangle + A_1B_0|\alpha,-\alpha,\alpha\rangle + A_1B_1|\alpha,-\alpha,-\alpha\rangle + A_2B_0|-\alpha,\alpha,\alpha\rangle + A_2B_1|-\alpha,\alpha,-\alpha\rangle + A_3B_0|-\alpha,-\alpha,\alpha\rangle + A_3B_1|-\alpha,-\alpha,-\alpha\rangle)_{A,A',B} \otimes |\psi\rangle_{123456} \quad , \tag{5}$$

where, normalization constant, $N_T = N_C \times N$ and $N = N_{AA'} \times N_B$

Simultaneous ABQT protocol may be outlined in following steps:

**Step1**: Alice mixes quantum-information modes,$(A)$ *and* $(A')$ with quantum channel's modes 1&2 respectively, on a 'symmetric beam splitter with phase shifter' BPS-1 and BPS-2 of which output modes may be indexed as 7,8 and 9,10. Similarly, Bob mixes quantum-information mode, (B) on third BPS-3 of which output modes are 11 and 12 (see Figure.1).

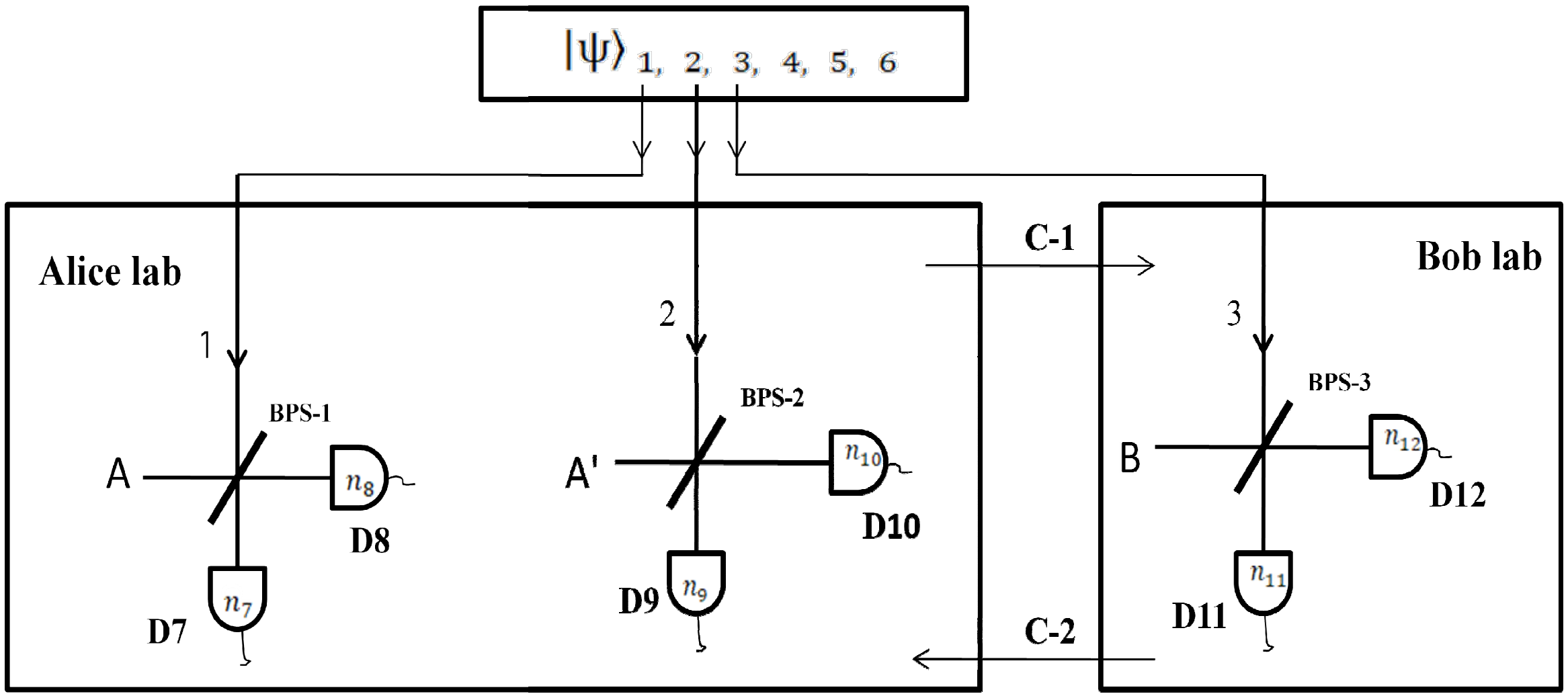

Figure1. Schematic diagram of the proposed scheme for ABQT. Alice mixes mode A,1 and $A'$,2 using (BPS-1) and (BPS-2) followed byphoton counting measurements at output modes $n_{7-10}$ by photon–number resolving detectors$D_{7-10}$,respectively. Simultaneously, Bob mixes mode B,3 on BPS-3 with output modes (11,12) using detector $D_{11-1}$ and communicate the outcomes (even and/or odd number of photons) to Alice and Bob using→classical channels $C_i$ (where i=1,2).

Applying Eq. (4) along with modes-mixing procedure spelled out in Step-1, Fig.1, we obtain global state, Eq. (A.1), in Appendix A Eq. (A.1), clearly, shows that there are null (vacuum) photons at various output modes of which presence play a significant role in photon counting measurements, see Step-3, Eqs. (7a-7h) below.

**Step2**- Alice performs photon-count measurements by the photon-number resolving detectors $n_{7,8}$ for output modes (7,8) from BPS-1 and $n_{9,10}$ for output modes (9,10) from BPS-2. Similarly, Bob does the same measurements via detectors $n_{11,12}$ for output modes (11,12) from BPS-3. That is to say, Eq. (A.1) in Appendix (A), collapses to a heralded state vector

$$|\chi(n_7,n_8,n_9,n_{10},n_{11},n_{12})\rangle_{4,5,6} = \langle n_7,n_8,n_9,n_{10},n_{11},n_{12}|\psi\rangle_{7,8,9,10,11,12,4,5,6} \quad (6)$$

Now, Alice sends the results of photon-number measurements $n_{7-10}$ to Bob through classical channel and, at the same time, Bob communicates, classically, his photon-numbers $n_{11-12}$ to Alice.

**Step 3**-Depending upon the null photon counts at various detectors, overall, eight cases may be identified for accomplishing simultaneous faithful (unit fidelity) ABQT. These cases are:

**Case-(I)**: Now, Alice and Bob may find photons$n_7 = 0, n_8 \neq 0$ from detectors $D_{7,8}$; photons $n_9 = 0, n_{10} \neq 0$ from detectors $D_{9,10}$ and photons $n_{11} = 0, n_{12} \neq 0$ from detectors $D_{11,12}$, i.e.,$|\chi^1(n_7{=}0, n_8, n_9{=}0, n_{10}, n_{11}{=}0, n_{12})\rangle_{4,5,6} = \langle 0, n_8, 0, n_{10}, 0, n_{12}|\psi\rangle_{7,8,9,10,11,12,4,5,6}$,

$$\begin{aligned}|\chi^1(0,n_8,0,n_{10},0,n_{12})\rangle_{4,5,6} &= N(A_0B_0|\alpha,\alpha,\alpha\rangle + (-1)^{n_{12}}A_0B_0|\alpha,\alpha,-\alpha\rangle + (-1)^{n_{10}}A_1B_0|\alpha,-\alpha,\alpha\rangle + (-1)^{n_{10}+n_{12}}A_1B_1|\alpha,-\alpha,-\alpha\rangle \\ &+ (-1)^{n_8}A_2B_0|-\alpha,\alpha,\alpha\rangle + (-1)^{n_8+n_{12}}A_2B_1|-\alpha,\alpha,-\alpha\rangle + (-1)^{n_8+n_{10}}A_3B_0|-\alpha,-\alpha,\alpha\rangle + (-1)^{n_8+n_{10}+n_{12}}A_3B_1|-\alpha,-\alpha,-\alpha\rangle)_{4,5,6}\end{aligned} \quad (7a)$$

Evidently, if photon-counts, $n_8$, $n_{10}$, and $n_{12}$ are even-numbered, the proposed ABQT scheme is simultaneously faithful for, the heralded state vector yields, $N_{AA'}(A_0|\alpha,\alpha\rangle + A_1|\alpha,-\alpha\rangle + A_2|-\alpha,\alpha\rangle + A_3|-\alpha,-\alpha\rangle)_{4,5} \otimes N_B(B_0|\alpha\rangle + B_1|-\alpha\rangle)_6$, which is tensor product of 'teleported' arbitrary bi-modal entangled coherent states and Schrodinger Cat coherent-states at Bob and Alice, respectively, tabulated in fifth row of Table-1, Appendix B.

**Case-(II)**: Next, Alice and Bob may find photons $n_7 \neq 0, n_8 = 0$ from detectors $D_{7,8}$; photons $n_9 \neq 0, n_{10} = 0$ from detectors $D_{9,10}$ and photons $n_{11} \neq 0, n_{12} = 0$ from detectors $D_{11,12}$, i.e.,

$$|\chi^2(n_7, n_8{=}0, n_9, n_{10}{=}0, n_{11}, n_{12}=0)\rangle_{4,5,6} = \langle n_7,0,n_9,0,n_{11},0|\psi\rangle_{7,8,9,10,11,12,4,5,6},$$

$$\begin{aligned}|\chi^2(n_7,0,n_9,0,n_{11},0)\rangle_{4,5,6} &= N(A_0B_0|-\alpha,-\alpha,-\alpha\rangle + (-1)^{n_{11}}A_0B_1|-\alpha,-\alpha,\alpha\rangle + (-1)^{n_9}A_1B_0|-\alpha,\alpha,-\alpha\rangle + (-1)^{n_9+n_{11}}A_1B_1|-\alpha,\alpha,\alpha\rangle \\ &+ (-1)^{n_7}A_2B_0|\alpha,-\alpha,-\alpha\rangle + (-1)^{n_7+n_{11}}A_2B_1|\alpha,-\alpha,\alpha\rangle + (-1)^{n_7+n_9}A_3B_0|\alpha,\alpha,-\alpha\rangle + (-1)^{n_7+n_9+n_{11}}A_3B_1|\alpha,\alpha,\alpha\rangle)_{4,5,6}\end{aligned} \quad (7b)$$

Eq. (7b) suggests that if $n_7, n_9$ and $n_{11}$ are even, $\left|\chi^1\right\rangle_{4,5,6} = P_{4,5,6}(\pi)\left|\chi^2\right\rangle_{4,5,6}$, where, P is $\psi$ - phase shifting unitary operator, $P(\psi)=\exp(-ia^{\dagger}a\psi)$ such that, $P(\psi)|\alpha\rangle \rightarrow |e^{i\psi}\alpha\rangle$, $taking \psi = \pi,\ |\alpha\rangle = |-\alpha\rangle$. This implies that simultaneous ABQT is faithfully achieved.

**Case-(III)**: Next, Alice and Bob may find photons $n_7 \neq 0, n_8 = 0$ from detectors $D_{7,8}$; photons $n_9 \neq 0, n_{10} = 0$ from detectors $D_{9,10}$ and photons $n_{11} = 0, n_{12} \neq 0$ from detectors $D_{11,12}$, i.e.,

$$\left|\chi^3\right\rangle_{4,5,6} = \left\langle n_7, 0, n_9, 0, 0, n_{12} \right|\psi\rangle_{7,8,9,10,11,12,4,5,6}$$

$$\left|\chi^3(n_7,0,n_9,0,0,n_{12})\right\rangle_{4,5,6} = N(A_0B_0|-\alpha,-\alpha,\alpha\rangle + (-1)^{n_{12}} A_0B_1|-\alpha,-\alpha,-\alpha\rangle + (-1)^{n_9} A_1B_0|-\alpha,\alpha,\alpha\rangle + (-1)^{n_9+n_{12}} A_1B_1|-\alpha,\alpha,-\alpha\rangle$$
$$+(-1)^{n_7} A_2B_0|\alpha,-\alpha,\alpha\rangle + (-1)^{n_7+n_{12}} A_2B_1|\alpha,-\alpha,-\alpha\rangle + (-1)^{n_7+n_9} A_3B_0|\alpha,\alpha,\alpha\rangle + (-1)^{n_7+n_9+n_{12}} A_3B_1|\alpha,\alpha,-\alpha\rangle)_{4,5,6} \quad (7c)$$

if $n_7, n_9$ and $n_{12}$ are even, $\left|\chi^1\right\rangle_{4,5,6} = P_{4,5}(\pi)\left|\chi^3\right\rangle_{4,5,6}$ i.e., simultaneous faithful ABQT.

**Case-(IV)**: Next, Alice and Bob may find photons $n_7 \neq 0, n_8 = 0$ from detectors $D_{7,8}$; photons $n_9 = 0, n_{10} \neq 0$ from detectors $D_{9,10}$ and photons $n_{11} \neq 0, n_{12} = 0$ from detectors $D_{11,12}$, i.e.,

$$\left|\chi^4\right\rangle_{4,5,6} = \left\langle n_7, 0, 0, n_{10}, n_{11}, 0 \right|\psi\rangle_{7,8,9,10,11,12,4,5,6}$$

$$\left|\chi^4(n_7,0,0,n_{10},n_{11},0)\right\rangle_{4,5,6} = N(A_0B_0|-\alpha,\alpha,-\alpha\rangle + (-1)^{n_{11}} A_0B_1|-\alpha,\alpha,\alpha\rangle + (-1)^{n_{10}} A_1B_0|-\alpha,-\alpha,-\alpha\rangle + (-1)^{n_{10}+n_{11}} A_1B_1|-\alpha,-\alpha,\alpha\rangle$$
$$+(-1)^{n_7} A_2B_0|\alpha,\alpha,-\alpha\rangle + (-1)^{n_7+n_{11}} A_2B_1|\alpha,\alpha,\alpha\rangle + (-1)^{n_7+n_{10}} A_3B_0|\alpha,-\alpha,-\alpha\rangle + (-1)^{n_7+n_{10}+n_{11}} A_3B_1|\alpha,-\alpha,\alpha\rangle)_{4,5,6} \quad (7d)$$

if $n_7, n_{10}$ and $n_{11}$ are even, $\left|\chi^1\right\rangle_{4,5,6} = P_{4,6}(\pi)\left|\chi^4\right\rangle_{4,5,6}$ i.e., simultaneous faithful ABQT.

**Case-(V)**: Next, Alice and Bob may find photons $n_7 \neq 0, n_8 = 0$ from detectors $D_{7,8}$; photons $n_9 = 0, n_{10} \neq 0$ from detectors $D_{9,10}$ and photons $n_{11} = 0, n_{12} \neq 0$ from detectors $D_{11,12}$, i.e.,

$$\left|\chi^5\right\rangle_{4,5,6} = \left\langle n_7, 0, 0, n_{10}, 0, n_{12} \right|\psi\rangle_{7,8,9,10,11,12,4,5,6}$$

$$\left|\chi^5(n_7,0,0,n_{10},0,n_{12})\right\rangle_{4,5,6} = N(A_0B_0|-\alpha,\alpha,\alpha\rangle + (-1)^{n_{12}} A_0B_1|-\alpha,\alpha,-\alpha\rangle + (-1)^{n_{10}} A_1B_0|-\alpha,-\alpha,\alpha\rangle + (-1)^{n_{10}+n_{12}} A_1B_1|-\alpha,-\alpha,-\alpha\rangle$$
$$+(-1)^{n_7} A_2B_0|\alpha,\alpha,\alpha\rangle + (-1)^{n_7+n_{12}} A_2B_1|\alpha,\alpha,-\alpha\rangle + (-1)^{n_7+n_{10}} A_3B_0|\alpha,-\alpha,\alpha\rangle + (-1)^{n_7+n_{10}+n_{12}} A_3B_1|\alpha,-\alpha,-\alpha\rangle)_{4,5,6} \quad (7e)$$

if $n_7, n_{10}$ and $n_{12}$ are even, $\left|\chi^1\right\rangle_{4,5,6} = P_4(\pi)\left|\chi^5\right\rangle_{4,5,6}$ i.e., simultaneous faithful ABQT.

**Case-(VI)**: Next, Alice and Bob may find photons $n_7 = 0, n_8 \neq 0$ from detectors $D_{7,8}$; photons $n_9 \neq 0, n_{10} = 0$ from detectors $D_{9,10}$ and photons $n_{11} \neq 0, n_{12} = 0$ from detectors $D_{11,12}$, i.e.,

$$\left|\chi^6\right\rangle_{4,5,6} = \left\langle 0, n_8, n_9, 0, n_{11}, 0 \right|\psi\rangle_{7,8,9,10,11,12,4,5,6}$$

$$\left|\chi^6(0,n_8,n_9,0,n_{11},0)\right\rangle_{4,5,6} = N(A_0B_0\left|\alpha,-\alpha,-\alpha\right\rangle + (-1)^{n_{11}} A_0B_1\left|\alpha,-\alpha,\alpha\right\rangle + (-1)^{n_9} A_1B_0\left|\alpha,\alpha,-\alpha\right\rangle + (-1)^{n_9+n_{11}} A_1B_1\left|\alpha,\alpha,\alpha\right\rangle$$
$$+(-1)^{n_8} A_2B_0\left|-\alpha,-\alpha,-\alpha\right\rangle + (-1)^{n_8+n_{11}} A_2B_1\left|-\alpha,-\alpha,\alpha\right\rangle + (-1)^{n_8+n_9} A_3B_0\left|-\alpha,\alpha,-\alpha\right\rangle + (-1)^{n_8+n_9+n_{11}} A_3B_1\left|-\alpha,\alpha,\alpha\right\rangle)_{4,5,6} \quad (7f)$$

if $n_7, n_9$ *and* $n_{11}$ are even, $\left|\chi^1\right\rangle_{4,5,6} = P_{5,6}(\pi)\left|\chi^6\right\rangle_{4,5,6}$ i.e., simultaneous faithful ABQT.

**Case-(VII)**: Next, Alice and Bob may find photons $n_7 = 0, n_8 \neq 0$ from detectors $D_{7,8}$; photons $n_9 \neq 0, n_{10} = 0$ from detectors $D_{9,10}$and photons $n_{11} = 0, n_{12} \neq 0$ from detectors $D_{11,12}$, i.e.,

$$\left|\chi^7\right\rangle_{4,5,6} = \left\langle 0, n_8, n_9, 0, 0, n_{12}\right|\psi\rangle_{7,8,9,10,11,12,4,5,6}$$

$$\left|\chi^7(0,n_8,n_9,0,0,n_{12})\right\rangle_{4,5,6} = N(A_0B_0\left|\alpha,-\alpha,\alpha\right\rangle + (-1)^{n_{12}} A_0B_1\left|\alpha,-\alpha,-\alpha\right\rangle + (-1)^{n_9} A_1B_0\left|\alpha,\alpha,\alpha\right\rangle + (-1)^{n_9+n_{12}} A_1B_1\left|\alpha,\alpha,-\alpha\right\rangle$$
$$+(-1)^{n_8} A_2B_0\left|-\alpha,-\alpha,\alpha\right\rangle + (-1)^{n_8+n_{12}} A_2B_1\left|-\alpha,-\alpha,-\alpha\right\rangle + (-1)^{n_8+n_9} A_3B_0\left|-\alpha,\alpha,\alpha\right\rangle + (-1)^{n_8+n_9+n_{12}} A_3B_1\left|-\alpha,\alpha,-\alpha\right\rangle)_{4,5,6} \quad (7g)$$

It can be observed that $n_8, n_9$ and $n_{12}$ are even, $\left|\chi^1\right\rangle_{4,5,6} = P_5(\pi)\left|\chi^7\right\rangle_{4,5,6}$ i.e., simultaneous faithful ABQT.

**Case-(VIII)**: Next, Alice and Bob may find photons $n_7 = 0, n_8 \neq 0$ from detectors $D_{7,8}$;$n_9 = 0, n_{10} \neq 0$ from detectors $D_{9,10}$ and photons $n_{11} \neq 0, n_{12} = 0$ from detectors $D_{11,12}$ i.e.,

$$\left|\chi^8\right\rangle_{4,5,6} = \left\langle 0, n_8, 0, n_{10}, n_{11}, 0\right|\psi\rangle_{7,8,9,10,11,12,4,5,6}$$

$$\left|\chi^8(0,n_8,0,n_{10},n_{11},0)\right\rangle_{4,5,6} = N(A_0B_0\left|\alpha,\alpha,-\alpha\right\rangle + (-1)^{n_{11}} A_0B_1\left|\alpha,\alpha,\alpha\right\rangle + (-1)^{n_{10}} A_1B_0\left|\alpha,-\alpha,-\alpha\right\rangle + (-1)^{n_{10}+n_{11}} A_1B_1\left|\alpha,-\alpha,\alpha\right\rangle$$
$$+(-1)^{n_8} A_2B_0\left|-\alpha,\alpha,-\alpha\right\rangle + (-1)^{n_8+n_{11}} A_2B_1\left|-\alpha,\alpha,\alpha\right\rangle + (-1)^{n_8+n_{10}} A_3B_0\left|-\alpha,-\alpha,-\alpha\right\rangle + (-1)^{n_8+n_{10}+n_{11}} A_3B_1\left|-\alpha,-\alpha,\alpha\right\rangle)_{4,5,6} \quad (7h)$$

Similarly, if $n_8, n_{10}$ and $n_{11}$ are even, $\left|\chi^1\right\rangle_{4,5,6} = P_6(\pi)\left|\chi^8\right\rangle_{4,5,6}$ i.e., simultaneous faithful ABQT.

In general, probability of successful accomplishment of simultaneousfaithful ABQT is nothing but finding number of photons $n_{7-12}$ in various modes (7-12),

$$P = (n_7, n_8, n_9, n_{10}, n_{11}, n_{12}) = \sum_{n_7, n_8, \ldots n_{12}} \left|\left\langle n_7, n_8, n_9, n_{10}, n_{11}, n_{12}\right|\psi\rangle_{7,8,9,10,11,12,4,5,6}\right|^2 \quad , \quad (8)$$

where, $\left|\psi\right\rangle_{7,8,9,10,11,12,4,5,6}$ is given in Eq. (A1) in Appendix A. Evaluation of Eq. (8) for Case-I Eq. (7a) is $P_{\chi^1} = (n_8, n_{10}, n_{12}, 0,0,0) = 1/8$ for even $n_8, n_{10}, n_{12}$. The total 'probability of success' ofsimultaneous faithful ABQT (denoted by $A \Leftrightarrow B$) scheme would, therefore, be $P_T^{A\Leftrightarrow B} = \sum_{i=1}^{8} P_{\chi^i} = 1$ for intense optical coherent states ($\alpha^2 \to \infty$). Evidently, total probability ($P_T^{A\Leftrightarrow B}$) is independent of α, the modal amplitude of optical coherent field as well as the unknown probability amplitudes $A_{0\text{-}3}$ of Alice's information-state and those of $B_{0,1}$ of Bob's information-state.

**Step 4**-Notably, Eqs. (7a-7h) do not demonstrate all exhaustive possibilities of detection of number of photons $n_{7-12}$from detectors $D_{7-12}$but there may exist several detection-events which are tabulated along with their consequent heralded states and local unitary operations in Alice's and Bob's lab in Tables (1-8), in Appendix B.

### 3. Probability of Success and Fidelity for Near-faithful Partial ABQT

A curious inspection of Tables (1-8) in Appendix B displays that not all detection-events accomplishes faithful ABQT but there exists cases wherein at least one uni-directional QT (either $A \rightarrow B$ or $B \rightarrow A$ or $A \rightleftharpoons B$). is near-faithful with finite probability of success, which may be termed as near-faithful partial ABQT, described below.

Firstly, Table-1, Eq. (7a), First row (Odd, Odd, Odd).

$$\left|\chi^1\right\rangle_{45} \otimes \left|\chi^1\right\rangle_6 = N_{AA'}(A_0|\alpha,\alpha\rangle - A_1|\alpha,-\alpha\rangle - A_2|-\alpha,\alpha\rangle + A_3|-\alpha,-\alpha\rangle)_{4,5} \otimes N_B(B_0|\alpha\rangle - B_1|-\alpha\rangle)_6 \tag{9}$$

Eq. (10), is not exactly same as information-states of Alice and Bob, but by application of displacement operator, $D_k(\delta) = \exp(\delta b_k^\dagger - \delta^* b_k)$ in Eq. (9) in Bob's lab, we obtain, taken, for simplicity, parameterization of probability amplitudes: $A_0= \cos\theta$ , $A_1= \sin\theta$, $A_2= \cos\phi$, $A_3= \sin\phi$ , $B_0= \cos\theta_1$ , $B_1= \sin\theta_1$,

$$D_4\left(\frac{i\pi}{2\alpha}\right)D_5\left(\frac{i\pi}{2\alpha}\right)\left|\chi^1\right\rangle_{45} = |\xi\rangle_{4,5} = N_{1AA'}\, e^{i\pi}\left(A_0\left|\frac{i\pi}{2\alpha}+\alpha\right\rangle\left|\frac{i\pi}{2\alpha}+\alpha\right\rangle + A_1\left|\frac{i\pi}{2\alpha}+\alpha\right\rangle\left|\frac{i\pi}{2\alpha}-\alpha\right\rangle + A_2\left|\frac{i\pi}{2\alpha}-\alpha\right\rangle\left|\frac{i\pi}{2\alpha}+\alpha\right\rangle + A_4\left|\frac{i\pi}{2\alpha}-\alpha\right\rangle\left|\frac{i\pi}{2\alpha}-\alpha\right\rangle\right)_{4,5} \tag{10}$$

And by following same technique in Alice lab, one obtains,

$$D_6\left(\frac{i\pi}{2\alpha}\right)\left|\chi^1\right\rangle_6 = |\zeta\rangle_6 = N_{1B}\, e^{i\pi/2}\left(B_0\left|\frac{i\pi}{2\alpha}+\alpha\right\rangle + B_1\left|\frac{i\pi}{2\alpha}-\alpha\right\rangle\right)_6 \tag{11}$$

Hence, the fidelities and probabilities for success of near-faithful ABQT in this detection-events, Table-1, Eq. (7a), First row (Odd, Odd, Odd), in Bob's lab $F_1^{(I)A\rightarrow B} = \left|{}_{AA'}\langle\psi|\xi\rangle_{4,5}\right|^2$ and in Alice's lab $F_1^{(I)B\rightarrow A} = \left|{}_B\langle\psi|\zeta\rangle_6\right|^2$ are given by, after insertion of Eqs. (10-11),

$$F_1^{(I)A\to B} = \left(N_{1AA'}N_{AA'}\right)^2 e^{\frac{-\pi^2}{4\alpha^2}}\left(\sum_{i=0}^{3} A_i^2 + 2e^{-4\alpha^2}\left(A_1A_2 - A_0A_3\right)\right)^2, P_1^{(I)A\to B} = \left(\frac{N_{AA'}}{8N_{1AA'}}\right)^2\left(\frac{1-e^{-2\alpha^2}}{1+e^{-2\alpha^2}}\right)^2 \quad (12)$$

$$F_1^{(I)B\to A} = \left(N_{1B}N_B\right)^2 e^{\frac{-\pi^2}{8\alpha^2}}\left(\sum_{i=0}^{1} B_i^2 + 2e^{-2\alpha^2}\left(B_0B_1\right)\right)^2, P_1^{(I)B\to A} = \left(\frac{N_B}{8N_{1B}}\right)^2 \quad (13)$$

where, $(\mathrm{N}_{1\mathrm{AA'}})^{-2} = \sum_{\mathrm{i}=0}^{3}\mathrm{A}_\mathrm{i}^2 - 2\mathrm{e}^{-2\alpha^2}(\mathrm{A}_0\mathrm{A}_1 + \mathrm{A}_0\mathrm{A}_2 + \mathrm{A}_1\mathrm{A}_3 + \mathrm{A}_2\mathrm{A}_3) + 2\mathrm{e}^{-4\alpha^2}(\mathrm{A}_0\mathrm{A}_3 + \mathrm{A}_1\mathrm{A}_2)$, and

$$(\mathrm{N}_{1\mathrm{B}})^{-2} = \sum_{\mathrm{j}=0}^{1}\mathrm{B}_\mathrm{j}^2 - 2\mathrm{e}^{-2\alpha^2}(\mathrm{B}_0\mathrm{B}_1).$$

One may note that a characteristic property of displacement operator, $D_I(\beta)|\delta\rangle_I = \exp\left[\frac{1}{2}\left(\beta\delta^* - \beta^*\delta\right)\right]|\beta+\delta\rangle_I$, has been applied along with the expression for scalar product of optical coherent states, $\langle\beta|\delta\rangle = \exp\left[-\left(|\beta|^2 + |\delta|^2 - 2\beta^*\delta\right)/2\right]$.Clearly, one may note from Eqs (12-13) that in both cases $(A\to B \quad and \quad B\to A)$ the teleportation is nearly faithful and thereby, 'Near-Faithfull Partial ABQT', plotted in Figure 2 to assess dependencies of of fidelities on probability-amplitudes $\theta,\phi$ as well $\alpha$ as onthe modal-amplitudes of optical coherent states. It is instructive to infer that Fidelities in Alice's and Bob's lab approaches, asymptotically, near unit-values for intense coherent optical field. Numerically, it may be seen that fidelities tends to unity for $\alpha \geq 5$.

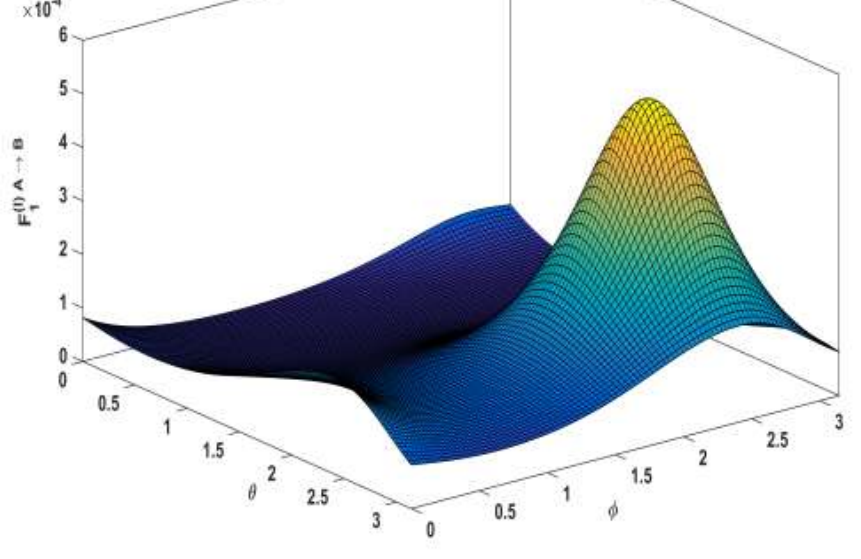

(a)

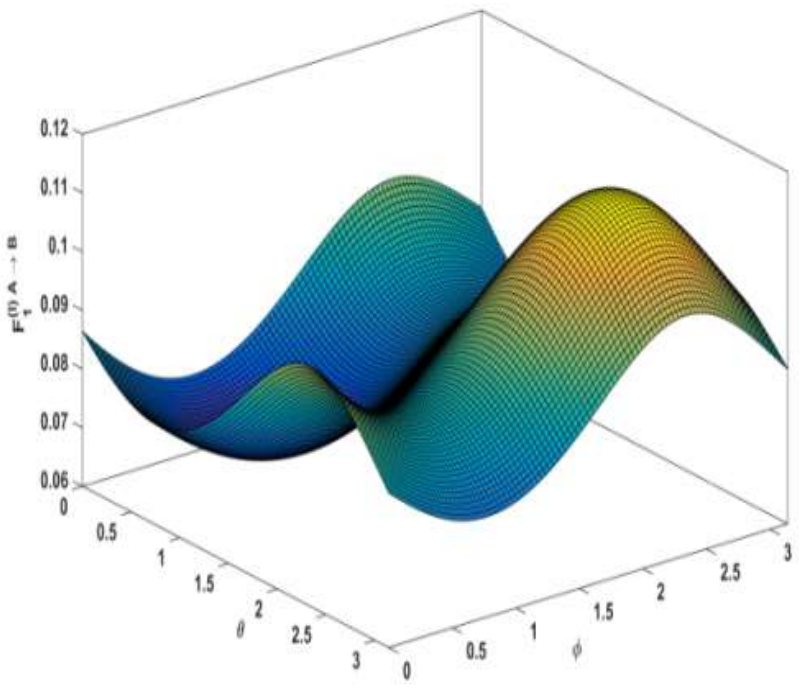

(b)

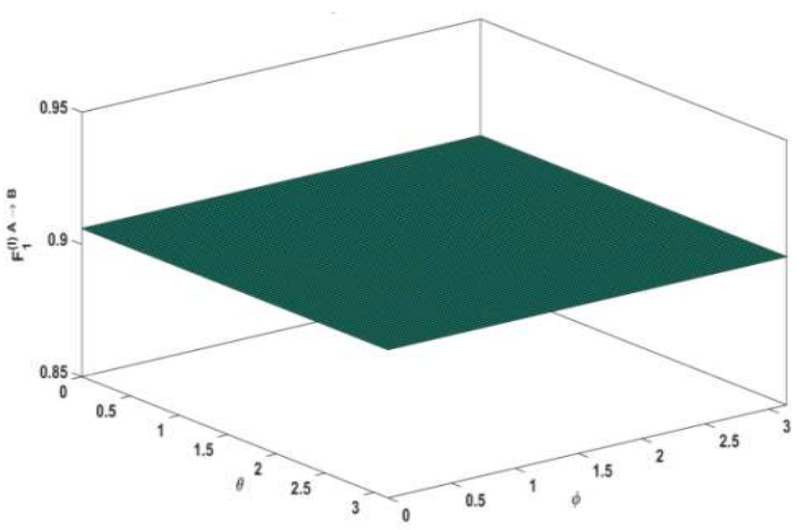

(b)

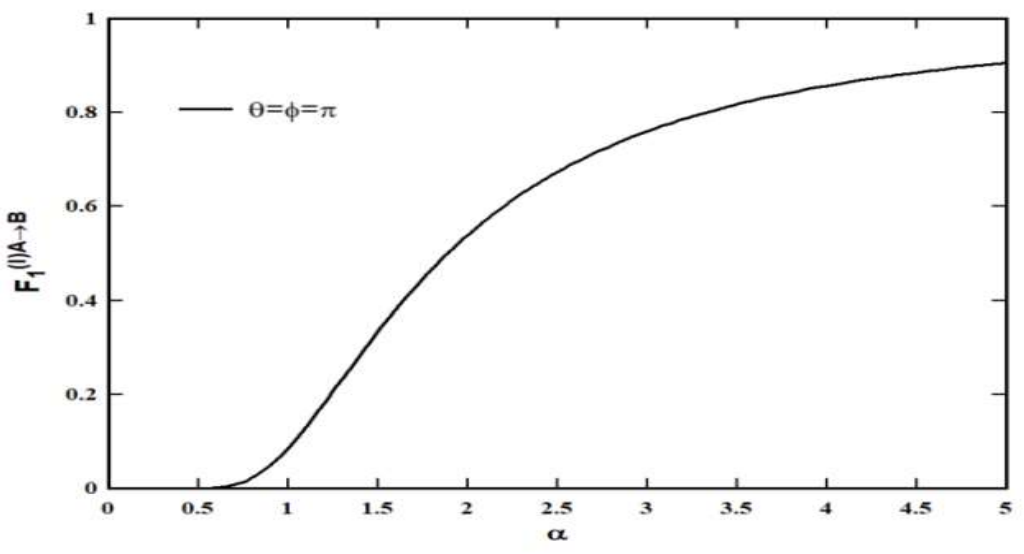

(d)

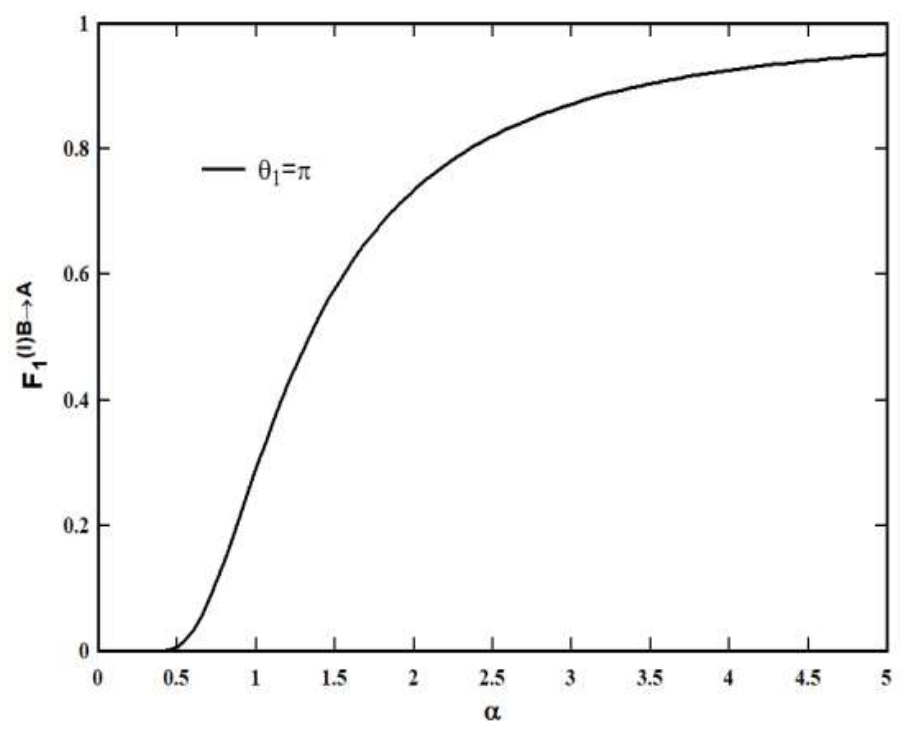

(e)

Figure 2. (a), (b) and (c) are $F_1^{(1)A\to B}$ Vs Θ, Φ for $\alpha = 0.5$, 1 and 5, respectively, (d) and (e) is the dependence of $F_1^{(I)A\to B}$ and $F_1^{(I)B\to A}$ on $\alpha$ for $\Theta_1=\pi$.

Following similar lineage one may evaluate fidelities and probabilities for other detection-events in Table-1, for, Eq. (7a), Second row (Odd, Odd, Even),

$$F_2^{(I)A\to B} = \left(N_{2AA'}N_{AA'}\right)^2 e^{\frac{-\pi^2}{4\alpha^2}}\left(\sum_{i=0}^{3} A_i^2 + 2e^{-2\alpha^2}\left(A_1A_2 - A_0A_3\right)\right)^2, P_2^{(I)A\to B} = \left(\frac{N_{AA'}}{8N_{2AA'}}\right)^2\left(\frac{1-e^{-2\alpha^2}}{1+e^{-2\alpha^2}}\right)^2 \qquad (14)$$

$$F_2^{(I)B\to A} = 1, \quad P_2^{(I)B\to A} = 1/64 \qquad (15)$$

$$where, (N_{2AA'})^{-2} = \sum_{i=0}^{3} A_i^2 - 2e^{-2\alpha^2}(A_0A_1 + A_0A_2 + A_1A_3 + A_2A_3) + 2e^{-4\alpha^2}(A_0A_3 + A_1A_2), \quad and \quad (N_{2B})^{-2} = \sum_{j=0}^{1} B_j^2 + 2e^{-2\alpha^2}(B_0B_1).$$

Obviously, Eq. (14) the teleportation in one direction ($A \to B$) is near faithful QT and in backward direction ($B \to A$) is faithful QT i.e, it is near-faithful partial ABQT.

Next, for Table-1, Eq. (7a), Third row (Odd, Even, Odd)

$$F_3^{(I)A\to B} = \left(N_{3AA'}N_{AA'}\right)^2 e^{\frac{-\pi^2}{8\alpha^2}}\left(\sum_{i=0}^{3} A_i^2 + 2e^{-2\alpha^2}\left(A_0A_1 + A_2A_3\right)\right)^2, P_3^{(1)A\to B} = \left(\frac{N_{AA'}}{8N_{3AA'}}\right)^2\left(\frac{1+e^{-4\alpha^2}}{1-e^{-4\alpha^2}}\right) \qquad (16)$$

$$F_3^{(I)B\to A} = \left(N_{3B}N_B\right)^2 e^{\frac{-\pi^2}{8\alpha^2}}\left(\sum_{j=0}^{1} B_j^2 + 2e^{-2\alpha^2}\left(B_0B_1\right)\right)^2, P_3^{(I)B\to A} = \left(\frac{N_B}{8N_{3B}}\right)^2 \qquad (17)$$

$$\text{where,} (N_{3AA'})^{-2} = \sum_{i=0}^{3} A_i^2 + 2e^{-2\alpha^2}(A_0A_1 + A_2A_3 - A_0A_2 - A_1A_3) - 2e^{-4\alpha^2}(A_0A_3 + A_1A_2) \quad \text{and}$$

$$(N_{3B})^{-2} = \sum_{j=0}^{1} B_j^2 - 2e^{-2\alpha^2}(B_0 B_1).$$

Evidently, Eqs. (16-17) demonstrates that in both directions$(A \to B)$ and $(B \to A)$ the QT is nearly faithful and, thus, it is near-faithful partial ABQT. Plot of Eq.(16) are in Figure 3 to assess dependence of fidelities on probability-amplitudes parameters $\Theta$, $\Phi$ as well as $\alpha$ on amplitudes of optical coherent states.

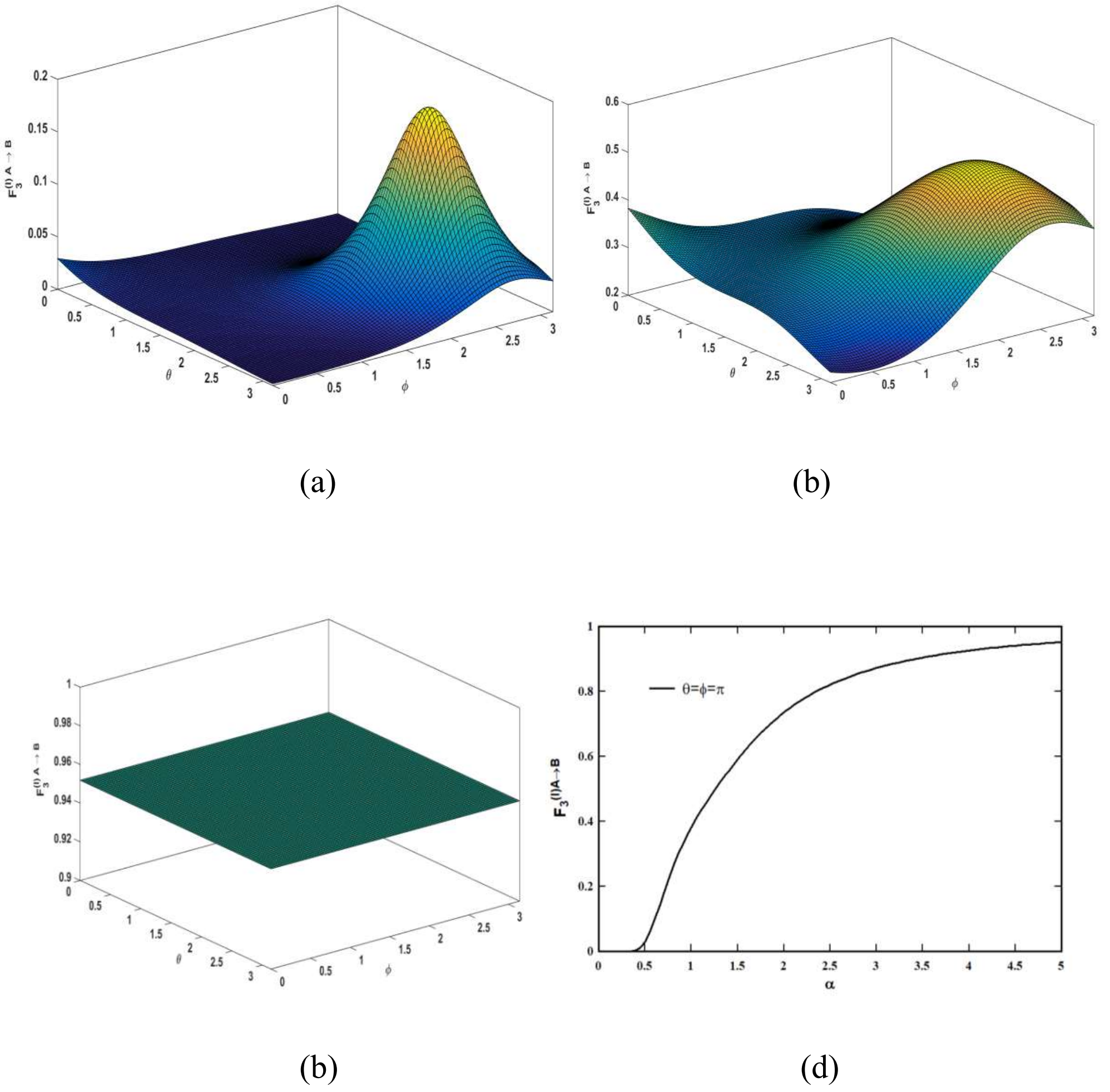

Figure 3. (a), (b) and (c) are $F_3^{(I)A\to B}$ vs $\Theta$, $\Phi$ for $\alpha = 0.5$, 1 and 5, respectively, (d) is the dependence of $F_3^{(I)A\to B}$ on $\alpha$ for $\Theta$=$\Phi$=$\pi$.

Next, for Table-1, Eq. (7a), Fourth row (Even, Odd, Odd),

$$F_4^{(I)A\to B} = \left(N_{4AA'} N_{AA'}\right)^2 e^{\frac{-\pi^2}{8\alpha^2}} \left( \sum_{i=0}^{3} A_i^2 + 2e^{-2\alpha^2}\left(A_0 A_2 + A_1 A_3\right) \right)^2, \; P_4^{(I)A\to B} = \left( \frac{N_{AA'}}{8 N_{4AA'}} \right)^2 \left( \frac{1+e^{-4\alpha^2}}{1-e^{-4\alpha^2}} \right) (18)$$

$$F_4^{(I)B\rightarrow A} = \left(N_{4B}N_B\right)^2 e^{\frac{-\pi^2}{8\alpha^2}}\left(\sum_{j=0}^{1} B_j^2 + 2e^{-2\alpha^2}\left(B_0B_1\right)\right)^2, P_4^{(I)B\rightarrow A} = \left(\frac{N_B}{8N_{4B}}\right)^2 \quad (19)$$

where, $(N_{4AA'})^{-2} = \sum_{i=0}^{3} A_i^2 - 2e^{-2\alpha^2}(A_0A_1 + A_0A_2 + A_1A_3 - A_2A_3) + 2e^{-4\alpha^2}(A_0A_3 + A_1A_2)$ and $(N_{4B})^{-2} = \sum_{j=0}^{1} B_j^2 - 2e^{-2\alpha^2}(B_0B_1)$.

Evidently, Eqs. (18-19) demonstrates that in both directions($A \rightarrow B$) and ($B \rightarrow A$) the QT is nearly faithful and, thus, it is near-faithful partial ABQT. Plot of Eqs. (18) are in Figure 4 to assess dependence of fidelities on probability-amplitudes parameters Θ, Φ, as well as $\alpha$ on amplitudes of optical coherent states.

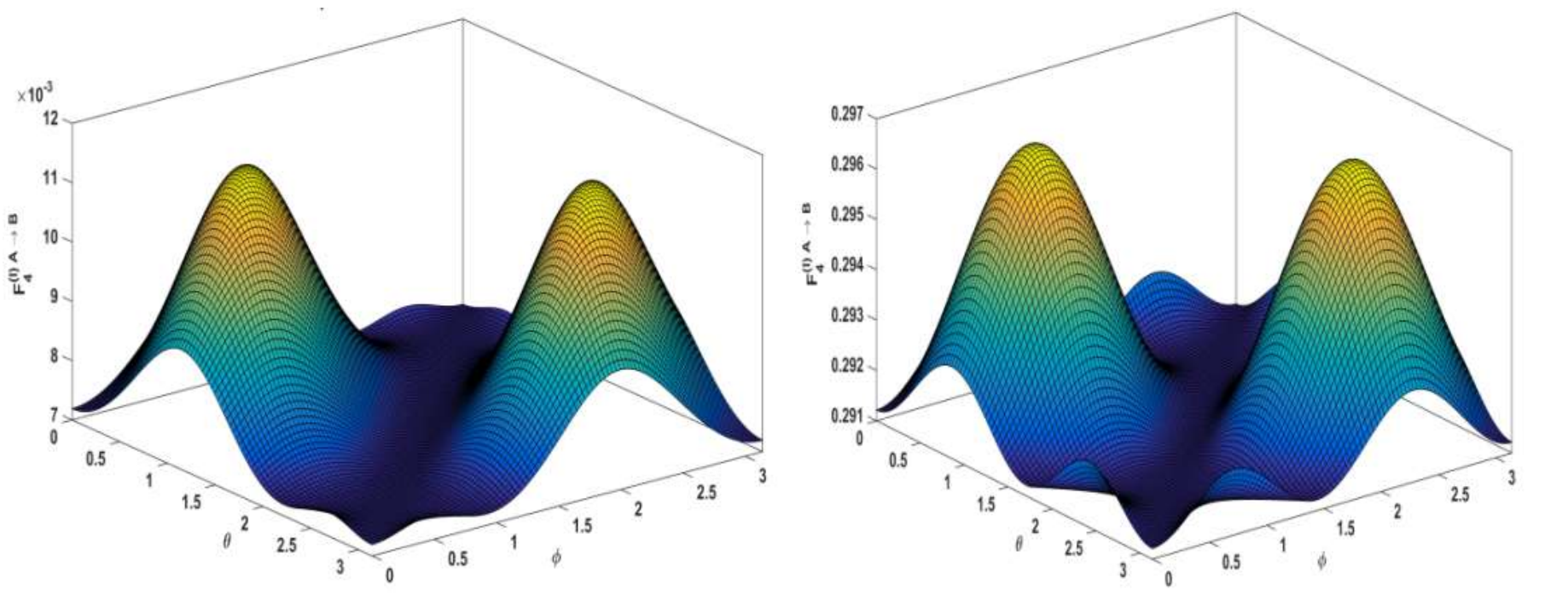

(a) (b)

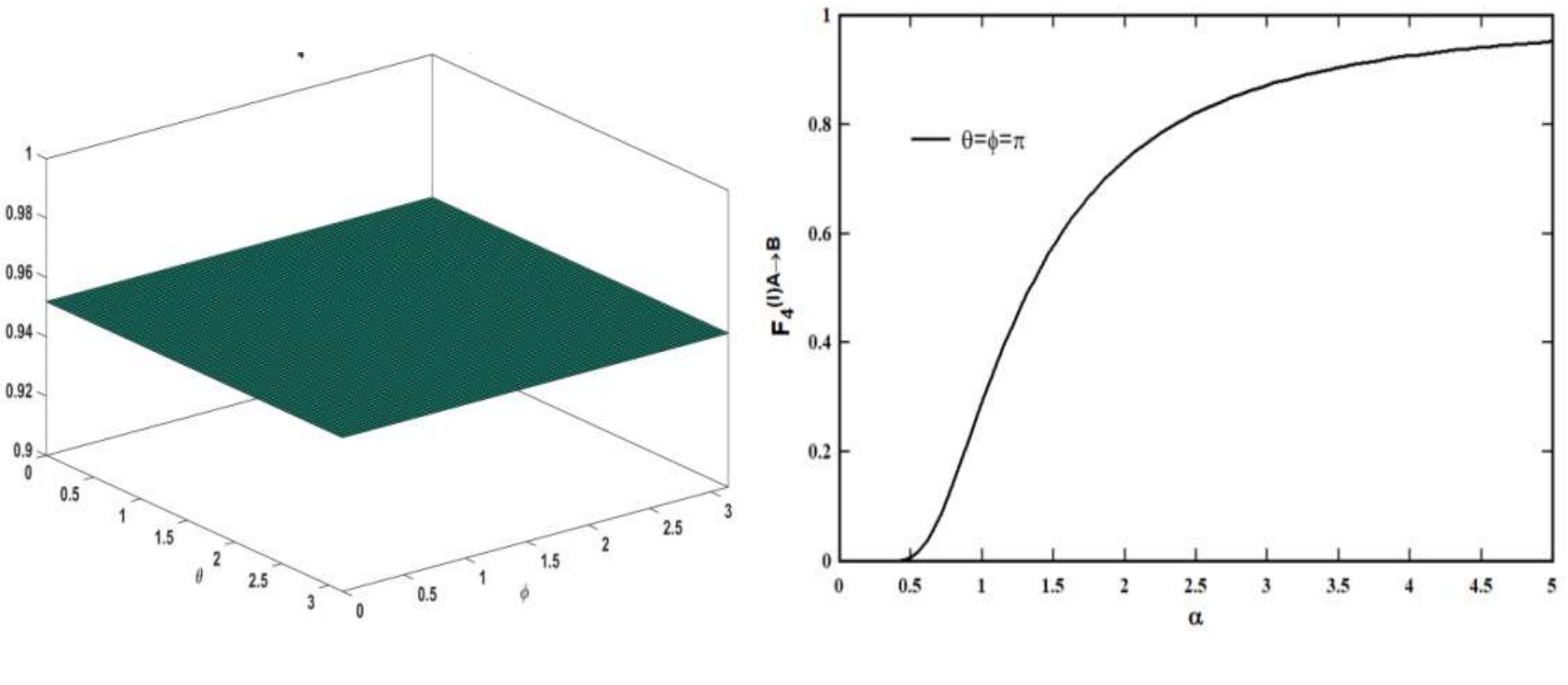

(c) (d)

Figure 4. (a), (b) and (c) are $F_4^{(I)A\rightarrow B}$ vs Θ, Φ for $\alpha$ = 0.5, 1 and 5, respectively, (d) is the dependence of $F_4^{(I)A\rightarrow B}$ on $\alpha$ for Θ=Φ=$\pi$.

Next, for Table-1, Eq. (7a), Fifth row (Even, Even, Even)

$$F_5^{(I)A\to B} = 1\ ,\ P_5^{(I)A\to B} = 1/64 \tag{20}$$

$$F_5^{(I)B\to A} = 1\ ,\ P_5^{(I)B\to A} = 1/64 \tag{21}$$

where, $(\mathrm{N}_{5\mathrm{AA}'})^{-2} = \sum_{i=0}^{3} \mathrm{A}_i^2 + 2e^{-2\alpha^2}(\mathrm{A}_0\mathrm{A}_2 + \mathrm{A}_1\mathrm{A}_3 + \mathrm{A}_0\mathrm{A}_1 + \mathrm{A}_2\mathrm{A}_3) + 2e^{-4\alpha^2}(\mathrm{A}_0\mathrm{A}_3 + \mathrm{A}_1\mathrm{A}_2)$ and

$(\mathrm{N}_{5\mathrm{B}})^{-2} = \sum_{j=0}^{1} \mathrm{B}_j^2 + 2e^{-2\alpha^2}(\mathrm{B}_0\mathrm{B}_1)$.

Clearly, in both directions $(A \to B)$ and $(B \to A)$ the faithful teleportation is achieved i.e, faithful partial ABQT described separately in forward and backward directions.

Next, Table-1, Eq. (7a), Six row (Even, Even, Odd),

$$F_6^{(I)A\to B} = 1, \quad P_6^{(I)A\to B} = 1/64 \tag{22}$$

$$F_6^{(I)B\to A} = (N_{6B}N_B)^2 e^{\frac{-\pi^2}{8\alpha^2}}\left(\sum_{j=0}^{1} B_j^2 + 2e^{-2\alpha^2}(B_0B_1)\right)^2, P_6^{(I)B\to A} = \left(\frac{N_B}{8N_{6B}}\right)^2 \tag{23}$$

where, $(N_{6AA'})^{-2} = \sum_{i=0}^{3} A_i^2 + 2e^{-2\alpha^2}(A_0A_2 + A_1A_3 + A_0A_1 + A_2A_3) + 2e^{-4\alpha^2}(A_0A_3 + A_1A_2)$ *and* $(N_{6B})^{-2} = \sum_{j=0}^{1} B_j^2 - 2e^{-2\alpha^2}(B_0B_1)$,

Evidently, Eq. (23) demonstrates that in one directions $(B \to A)$ the QT is nearly faithful and, thus, it is near-faithful partial ABQT.

Table-1, Eq. (7a), Seventh row (Even, Odd, Even),

$$F_7^{(1)B\to A} = 1 \quad and \quad P_7^{(1)B\to A} = 1/64 \tag{24}$$

$$F_7^{(I)A\to B} = (N_{7AA'}N_{AA'})^2 e^{\frac{-\pi^2}{8\alpha^2}}\left(\sum_{i=0}^{3} A_i^2 + 2e^{-2\alpha^2}(A_0A_2 + A_1A_3)\right)^2, P_7^{(I)A\to B} = \left(\frac{N_{AA'}}{8N_{7AA'}}\right)^2\left(\frac{1+e^{-4\alpha^2}}{1-e^{-4\alpha^2}}\right) \tag{25}$$

*where,* $(N_{7AA'})^{-2} = \sum_{i=0}^{3} A_i^2 - 2e^{-2\alpha^2}(A_0A_1 + A_0A_2 + A_1A_3 - A_2A_3) + 2e^{-4\alpha^2}(A_0A_3 + A_1A_2)$ and $(N_{7B})^{-2} = \sum_{j=0}^{1} B_j^2 - 2e^{-2\alpha^2}(B_0B_1)$.

Clearly, Eq. (25) demonstrates that in one directions $(A \to B)$ the QT is nearly faithful and, thus, it is near-faithful partial ABQT.

Next Table-1, Eq. (7a), Eighth row (Odd, Even, Even),

$$F_8^{(1)B\to A} = 1 \quad and \quad P_8^{(1)B\to A} = 1/64 \tag{26}$$

$$F_8^{(I)A\to B} = (N_{8AA'}N_{AA'})^2 e^{\frac{-\pi^2}{8\alpha^2}}\left(\sum_{i=0}^{3} A_i^2 + 2e^{-2\alpha^2}(A_0A_1 + A_2A_3)\right)^2, \quad P_8^{(I)A\to B} = \left(\frac{N_{AA'}}{8N_{8AA'}}\right)^2 \tag{27}$$

where, $(N_{8AA'})^{-2} = \sum_{i=0}^{3} A_i^2 + 2e^{-2\alpha^2}(A_0A_1 + A_2A_3 - A_0A_2 - A_1A_3) - 2e^{-4\alpha^2}(A_0A_3 + A_1A_2)$ and

$(N_{8B})^{-2} = \sum_{j=0}^{1} B_j^2 - 2e^{-2\alpha^2}(B_0B_1)$.

Clearly, Eq. (27) demonstrates that in one directions $(A \to B)$ the QT is nearly faithful and, thus, it is near-faithful partial ABQT.

One may calculate fidelities for entire Cases (I-VIII) described by Eqs. (7a-7h). It is seen that not all fidelities are different for,

$$F_1^{(I)A\to B} = F_1^{(II)A\to B} = F_1^{(III)A\to B} = F_1^{(IV)A\to B} = F_1^{(V)A\to B} = F_1^{(VI)A\to B} = F_1^{(VII)A\to B} = F_1^{(VIII)A\to B} = F_2^{(I)A\to B} = F_2^{(II)A\to B} = F_2^{(III)A\to B} = F_2^{(IV)A\to B} = F_2^{(V)A\to B} = F_2^{(VI)A\to B} = F_2^{(VII)A\to B} = F_2^{(VIII)A\to B},$$

$$F_3^{(I)A\to B} = F_3^{(II)A\to B} = F_3^{(III)A\to B} = F_3^{(IV)A\to B} = F_3^{(V)A\to B} = F_3^{(VI)A\to B} = F_3^{(VII)A\to B} = F_3^{(VIII)A\to B} = F_8^{(I)A\to B} = F_8^{(II)A\to B} = F_8^{(III)A\to B} = F_8^{(IV)A\to B} = F_8^{(V)A\to B} = F_8^{(VI)A\to B} = F_8^{(VII)A\to B} = F_8^{(VIII)A\to B},$$

$$F_4^{(I)A\to B} = F_4^{(II)A\to B} = F_4^{(III)A\to B} = F_4^{(IV)A\to B} = F_4^{(V)A\to B} = F_4^{(VI)A\to B} = F_4^{(VII)A\to B} = F_4^{(VIII)A\to B} = F_7^{(I)A\to B} = F_7^{(II)A\to B} = F_7^{(III)A\to B} = F_7^{(IV)A\to B} = F_7^{(V)A\to B} = F_7^{(VI)A\to B} = F_7^{(VII)A\to B} = F_7^{(VIII)A\to B},$$

$$F_5^{(I)A\to B} = F_5^{(II)A\to B} = F_5^{(III)A\to B} = F_5^{(IV)A\to B} = F_5^{(V)A\to B} = F_5^{(VI)A\to B} = F_5^{(VII)A\to B} = F_5^{(VIII)A\to B} F_6^{(I)A\to B} = F_6^{(II)A\to B} = F_6^{(III)A\to B} = F_6^{(IV)A\to B} = F_6^{(V)A\to B} = F_6^{(VI)A\to B} = F_6^{(VII)A\to B} = F_6^{(VIII)A\to B}$$

Similarly, one may notice that Fidelities from Bob to Alice bears following equalities,

$$F_1^{(I)B\to A} = F_1^{(II)B\to A} = F_1^{(III)B\to A} = F_1^{(IV)B\to A} = F_1^{(V)B\to A} = F_1^{(VI)B\to A} = F_1^{(VII)B\to A} = F_1^{(VIII)B\to A} = F_3^{(I)B\to A} = F_3^{(II)B\to A} = F_3^{(III)B\to A} = F_3^{(IV)B\to A} = F_3^{(V)B\to A} = F_3^{(VI)B\to A} = F_3^{(VII)B\to A} = F_3^{(VIII)B\to A} = F_4^{(I)B\to A} = F_4^{(II)B\to A} = F_4^{(III)B\to A} = F_4^{(IV)B\to A} = F_4^{(V)B\to A} = F_4^{(VI)B\to A} = F_4^{(VII)B\to A} + F_4^{(VIII)B\to A} = F_6^{(I)B\to A} = F_6^{(II)B\to A} = F_6^{(III)B\to A} = F_6^{(IV)B\to A} = F_6^{(V)B\to A} = F_6^{(VI)B\to A} = F_6^{(VII)B\to A} = F_6^{(VIII)B\to A},$$

$$F_2^{(I)B\to A} = F_2^{(II)B\to A} = F_2^{(III)B\to A} = F_2^{(IV)B\to A} = F_2^{(V)B\to A} = F_2^{(VI)B\to A} = F_2^{(VII)B\to A} = F_2^{(VIII)B\to A} = F_5^{(I)B\to A} = F_5^{(II)B\to A} = F_5^{(III)B\to A} = F_5^{(IV)B\to A} = F_5^{(V)B\to A} = F_5^{(VI)B\to A} = F_5^{(VII)B\to A} = F_5^{(VIII)B\to A} = F_7^{(I)B\to A} = F_7^{(II)B\to A} = F_7^{(III)B\to A} = F_7^{(IV)B\to A} = F_7^{(V)B\to A} = F_7^{(VI)B\to A} = F_7^{(VII)B\to A} = F_7^{(VIII)B\to A} = F_8^{(I)B\to A} = F_8^{(II)B\to A} = F_8^{(III)B\to A} = F_8^{(IV)B\to A} = F_8^{(V)B\to A} = F_8^{(VI)B\to A} = F_8^{(VII)B\to A} = F_8^{(VIII)B\to A}$$

Finally the 'Average Fidelity' of entire scheme from$(A \to B)$ and $(B \to A)$ can be defined as,

$$F_{av}^{A\to B}=\sum_{i=1}^{8}(F_i^{(I)A\to B}P_i^{(I)A\to B}+F_i^{(II)A\to B}P_i^{(II)A\to B}+.........+F_i^{(VIII)A\to B}P_i^{(VIII)A\to B}) \qquad (28)$$

$$F_{av}^{B\to A}=\sum_{j=1}^{8}(F_j^{(I)B\to A}P_i^{(I)B\to A}+F_j^{(II)B\to A}P_j^{(II)B\to A}+.........+F_j^{(VIII)B\to A}P_j^{(VIII)B\to A}) \qquad (29)$$

Plots of Eqs. (28-29), Figure 5 clearly, displays that Average fidelity is independent of probability-amplitudes parameters Θ, Φ as well as $\alpha$ on amplitudes of optical coherent states.

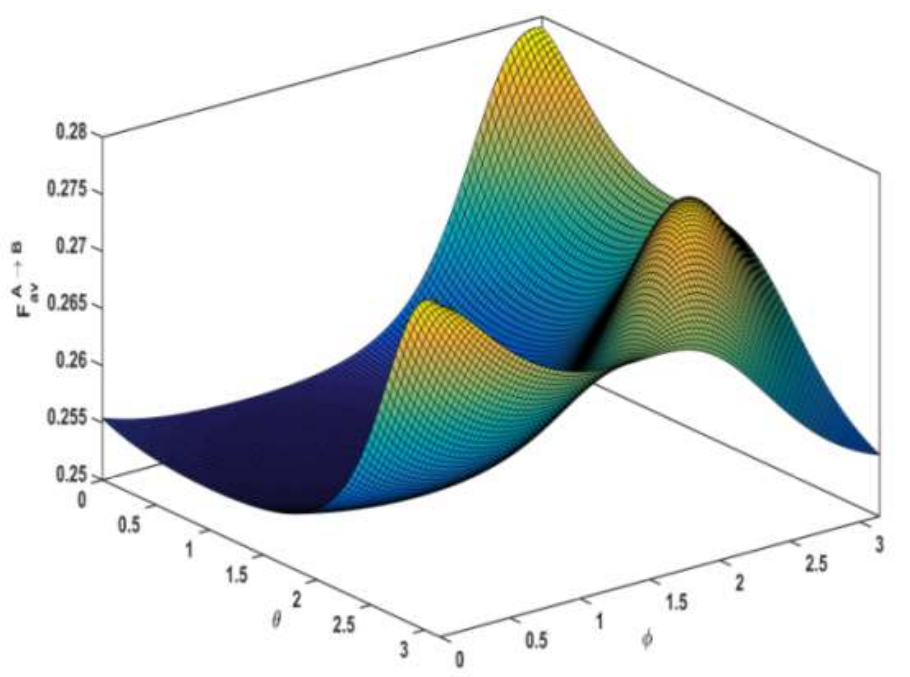

(a)

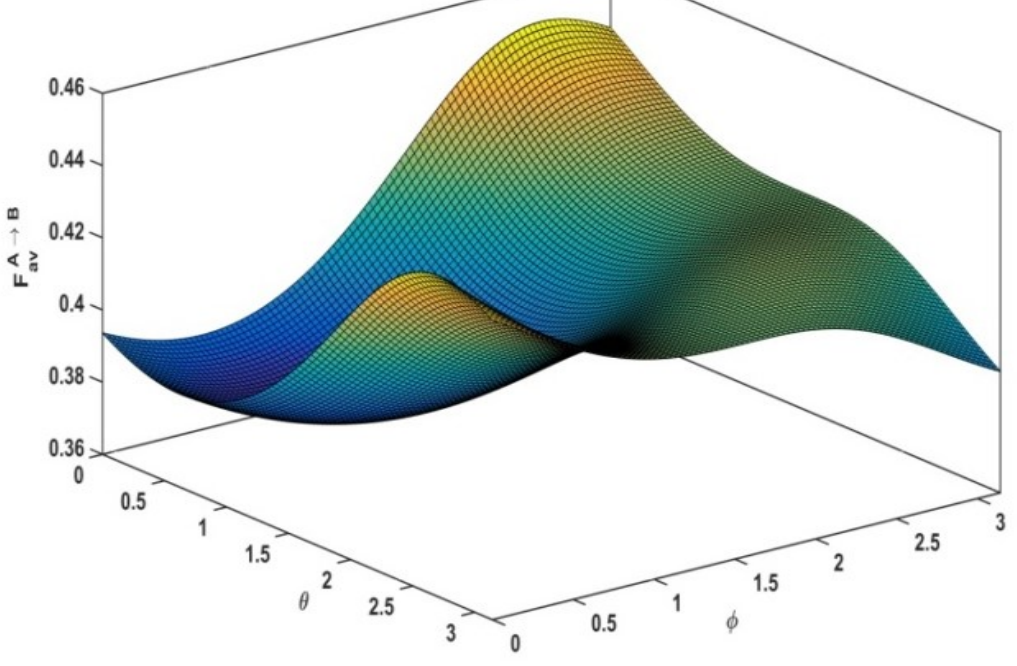

(b)

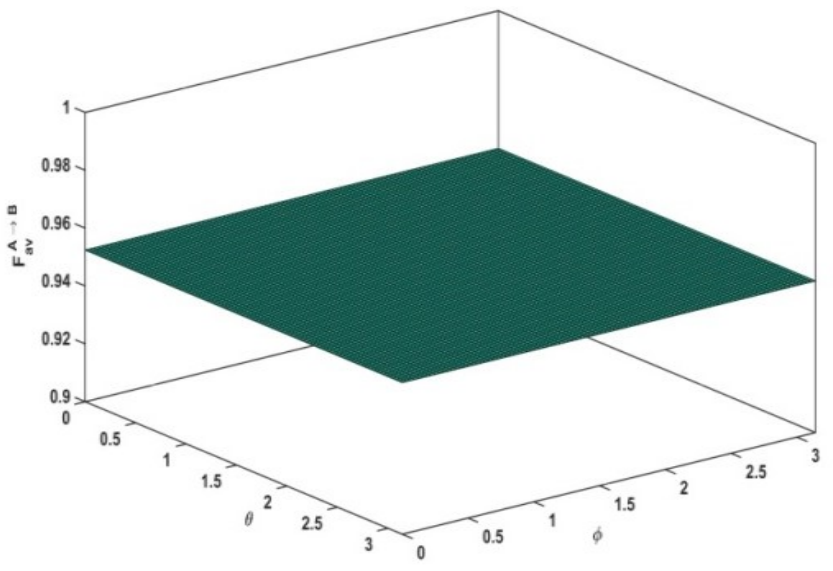

(c)

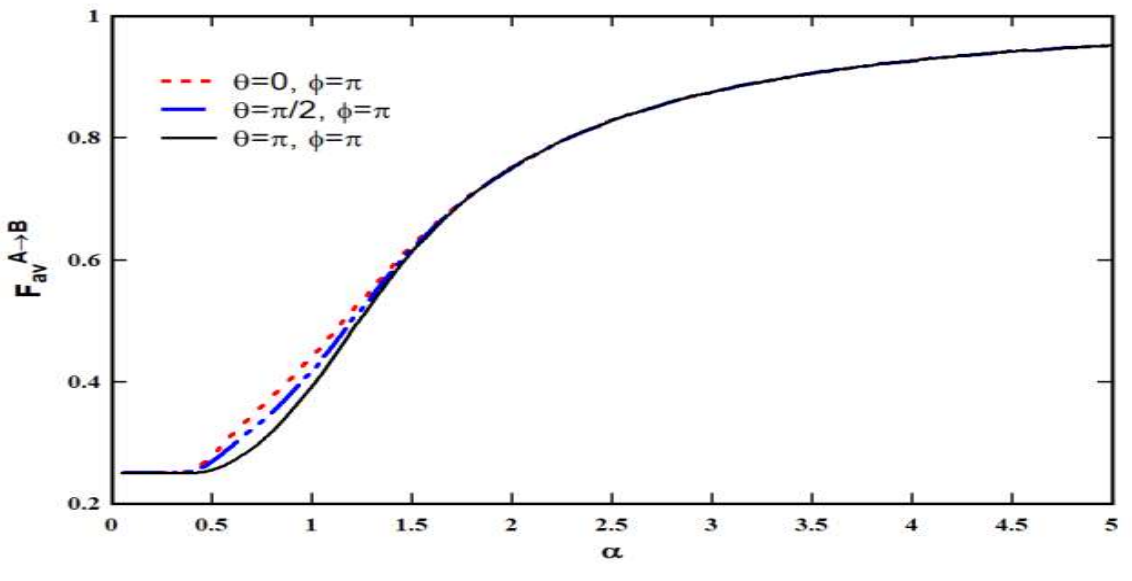

(d)

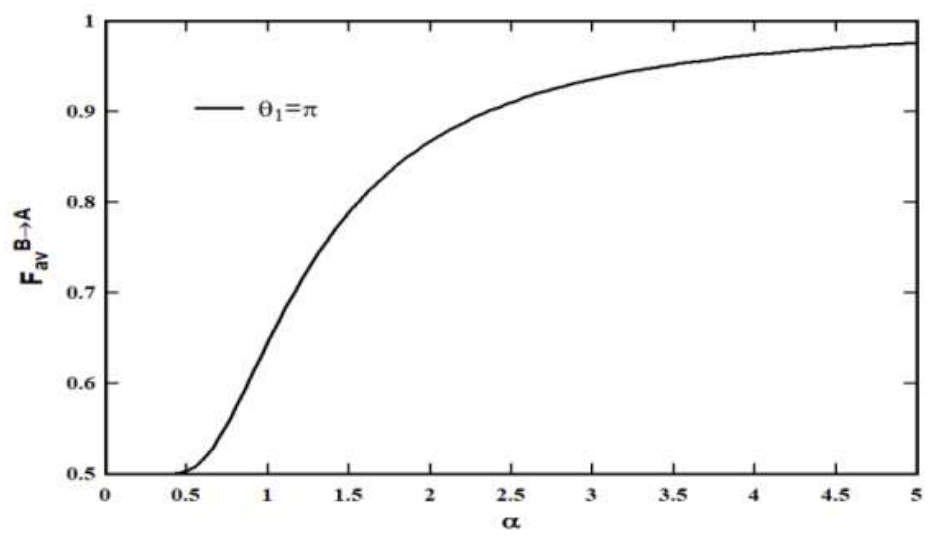

(e)

Figure 5. (a), (b) and (c) are $F_{av}^{A\to B}$ vs Θ, Φ for $\alpha$ = 0.5, 1 and 5 respectively, (d) is the dependence of $F_{av}^{A\to B}$ Vs [Θ,Φ]= [0, $\pi$], [Θ,Φ]= [$\pi/2$, $\pi$ ], [Θ,Φ]= [$\pi$, $\pi$]. and (e) $F_{av}^{B\to A}$ dependent on $\alpha$ for $\Theta_1=\pi$.

## Communication complexity and future prospects.

We have presented a protocol for a simultaneous faithful ABQT and near-faithful partial ABQT schemes. The communication complexity of the protocol may, easily, be recognized as three classical bits: two classical bit from Alice to Bob and one classical bit from Bob to Alice, if one encodes (ODD, EVEN) as (0,1), respectively, and the three e-bits because three Bell Coherent-states are needed to prepare quantum resource employed as the quantum channel. The scheme worked out, here, may be generalized by invoking hybrid QT in which QT is for forward direction and remote-state preparation in reverse direction or increasing number of parties involved in QT, which necessitates new publications.

## Acknowledgement

We acknowledge Prof. Lev Vaidman, School of Physics and Astronomy, Tel Aviv University, Israel and Prof. Mark M. Wilde, School of Electrical and Computer Engineering Cornell University, New York for drawing our attention toward their investigations on continuous variable quantum teleportation and bidirectional quantum teleportation.

**Appendix A**: Mixing various modes of quantum-information states (Eqs.1 and 2) and those of quantum channel (Eq.3) at 'Symmetric Beam Splitter with phase shifter', one obtains,

$$
\begin{aligned}
&|\phi\rangle_{7,8,9,10,11,12,4,5,6} = N_T[A_0B_0(|\sqrt{2\alpha},0\rangle_{7,8}|\sqrt{2\alpha},0\rangle_{9,10}|\sqrt{2\alpha},0\rangle_{11,12}|-\alpha,-\alpha,-\alpha\rangle_{4,5,6}+|\sqrt{2\alpha},0\rangle_{7,8}|\sqrt{2\alpha},0\rangle_{9,10}|0,\sqrt{2\alpha}\rangle_{11,12}|-\alpha,-\alpha,\alpha\rangle_{4,5,6}+\\
&|\sqrt{2\alpha},0\rangle_{7,8}|0,\sqrt{2\alpha}\rangle_{9,10}|\sqrt{2\alpha},0\rangle_{11,12}|-\alpha,\alpha,-\alpha\rangle_{4,5,6}+|\sqrt{2\alpha},0\rangle_{7,8}|0,\sqrt{2\alpha}\rangle_{9,10}|0,\sqrt{2\alpha}\rangle_{11,12}|-\alpha,\alpha,\alpha\rangle_{4,5,6}+\\
&|0,\sqrt{2\alpha}\rangle_{7,8}|\sqrt{2\alpha},0\rangle_{9,10}|\sqrt{2\alpha},0\rangle_{11,12}|\alpha,-\alpha,-\alpha\rangle_{4,5,6}+|0,\sqrt{2\alpha}\rangle_{7,8}|\sqrt{2\alpha},0\rangle_{9,10}|0,\sqrt{2\alpha}\rangle_{11,12}|\alpha,-\alpha,\alpha\rangle_{4,5,6}+\\
&|0,\sqrt{2\alpha}\rangle_{7,8}|0,\sqrt{2\alpha}\rangle_{9,10}|\sqrt{2\alpha},0\rangle_{11,12}|\alpha,\alpha,-\alpha\rangle_{4,5,6}+|0,\sqrt{2\alpha}\rangle_{7,8}|0,\sqrt{2\alpha}\rangle_{9,10}|0,\sqrt{2\alpha}\rangle_{11,12}|\alpha,\alpha,\alpha\rangle_{4,5,6})+\\
&A_0B_1(|\sqrt{2\alpha},0\rangle_{7,8}|\sqrt{2\alpha},0\rangle_{9,10}|0,-\sqrt{2\alpha}\rangle_{11,12}|-\alpha,-\alpha,-\alpha\rangle_{4,5,6}+|\sqrt{2\alpha},0\rangle_{7,8}|\sqrt{2\alpha},0\rangle_{9,10}|-\sqrt{2\alpha},0\rangle_{11,12}|-\alpha,-\alpha,\alpha\rangle_{4,5,6}+\\
&|\sqrt{2\alpha},0\rangle_{7,8}|0,\sqrt{2\alpha}\rangle_{9,10}|0,-\sqrt{2\alpha}\rangle_{11,12}|-\alpha,\alpha,-\alpha\rangle_{4,5,6}+|\sqrt{2\alpha},0\rangle_{7,8}|0,\sqrt{2\alpha}\rangle_{9,10}|-\sqrt{2\alpha},0\rangle_{11,12}|-\alpha,\alpha,\alpha\rangle_{4,5,6}+\\
&|0,\sqrt{2\alpha}\rangle_{7,8}|\sqrt{2\alpha},0\rangle_{9,10}|0,-\sqrt{2\alpha}\rangle_{11,12}|\alpha,-\alpha,-\alpha\rangle_{4,5,6}+|0,\sqrt{2\alpha}\rangle_{7,8}|\sqrt{2\alpha},0\rangle_{9,10}|-\sqrt{2\alpha},0\rangle_{11,12}|\alpha,-\alpha,\alpha\rangle_{4,5,6}+\\
&|0,\sqrt{2\alpha}\rangle_{7,8}|0,\sqrt{2\alpha}\rangle_{9,10}|0,-\sqrt{2\alpha}\rangle_{11,12}|\alpha,\alpha,-\alpha\rangle_{4,5,6}+|0,\sqrt{2\alpha}\rangle_{7,8}|0,\sqrt{2\alpha}\rangle_{9,10}|-\sqrt{2\alpha},0\rangle_{11,12}|\alpha,\alpha,\alpha\rangle_{4,5,6})+\\
&A_1B_0(|\sqrt{2\alpha},0\rangle_{7,8}|0,-\sqrt{2\alpha}\rangle_{9,10}|\sqrt{2\alpha},0\rangle_{11,12}|-\alpha,-\alpha,-\alpha\rangle_{4,5,6}+|\sqrt{2\alpha},0\rangle_{7,8}|0,-\sqrt{2\alpha}\rangle_{9,10}|0,\sqrt{2\alpha}\rangle_{11,12}|-\alpha,-\alpha,\alpha\rangle_{4,5,6}+\\
&|\sqrt{2\alpha},0\rangle_{7,8}|-\sqrt{2\alpha},0\rangle_{9,10}|\sqrt{2\alpha},0\rangle_{11,12}|-\alpha,\alpha,-\alpha\rangle_{4,5,6}+|\sqrt{2\alpha},0\rangle_{7,8}|-\sqrt{2\alpha},0\rangle_{9,10}|0,\sqrt{2\alpha}\rangle_{11,12}|-\alpha,\alpha,\alpha\rangle_{4,5,6}+\\
&|0,\sqrt{2\alpha}\rangle_{7,8}|0,-\sqrt{2\alpha}\rangle_{9,10}|\sqrt{2\alpha},0\rangle_{11,12}|\alpha,-\alpha,-\alpha\rangle_{4,5,6}+|0,\sqrt{2\alpha}\rangle_{7,8}|0,-\sqrt{2\alpha}\rangle_{9,10}|0,\sqrt{2\alpha}\rangle_{11,12}|\alpha,-\alpha,\alpha\rangle_{4,5,6}+\\
&|0,\sqrt{2\alpha}\rangle_{7,8}|-\sqrt{2\alpha},0\rangle_{9,10}|\sqrt{2\alpha},0\rangle_{11,12}|\alpha,\alpha,-\alpha\rangle_{4,5,6}+|0,\sqrt{2\alpha}\rangle_{7,8}|-\sqrt{2\alpha},0\rangle_{9,10}|0,\sqrt{2\alpha}\rangle_{11,12}|\alpha,\alpha,\alpha\rangle_{4,5,6})+\\
&A_1B_1(|\sqrt{2\alpha},0\rangle_{7,8}|0,-\sqrt{2\alpha}\rangle_{9,10}|0,-\sqrt{2\alpha}\rangle_{11,12}|-\alpha,-\alpha,-\alpha\rangle_{4,5,6}+|\sqrt{2\alpha},0\rangle_{7,8}|0,-\sqrt{2\alpha}\rangle_{9,10}|-\sqrt{2\alpha},0\rangle_{11,12}|-\alpha,-\alpha,\alpha\rangle_{4,5,6}+\\
&|\sqrt{2\alpha},0\rangle_{7,8}|-\sqrt{2\alpha},0\rangle_{9,10}|0,-\sqrt{2\alpha}\rangle_{11,12}|-\alpha,\alpha,-\alpha\rangle_{4,5,6}+|\sqrt{2\alpha},0\rangle_{7,8}|-\sqrt{2\alpha},0\rangle_{9,10}|-\sqrt{2\alpha},0\rangle_{11,12}|-\alpha,\alpha,\alpha\rangle_{4,5,6}+\\
&|0,\sqrt{2\alpha}\rangle_{7,8}|0,-\sqrt{2\alpha}\rangle_{9,10}|0,-\sqrt{2\alpha}\rangle_{11,12}|\alpha,-\alpha,-\alpha\rangle_{4,5,6}+|0,\sqrt{2\alpha}\rangle_{7,8}|0,-\sqrt{2\alpha}\rangle_{9,10}|-\sqrt{2\alpha},0\rangle_{11,12}|\alpha,-\alpha,\alpha\rangle_{4,5,6}+\\
&|0,\sqrt{2\alpha}\rangle_{7,8}|-\sqrt{2\alpha},0\rangle_{9,10}|0,-\sqrt{2\alpha}\rangle_{11,12}|\alpha,\alpha,-\alpha\rangle_{4,5,6}+|0,\sqrt{2\alpha}\rangle_{7,8}|-\sqrt{2\alpha},0\rangle_{9,10}|-\sqrt{2\alpha},0\rangle_{11,12}|\alpha,\alpha,\alpha\rangle_{4,5,6})+\\
&A_2B_0(|0,-\sqrt{2\alpha},\rangle_{7,8}|\sqrt{2\alpha},0\rangle_{9,10}|\sqrt{2\alpha},0\rangle_{11,12}|-\alpha,-\alpha,-\alpha\rangle_{4,5,6}+|0,-\sqrt{2\alpha}\rangle_{7,8}|\sqrt{2\alpha},0\rangle_{9,10}|0,\sqrt{2\alpha}\rangle_{11,12}|-\alpha,-\alpha,\alpha\rangle_{4,5,6}+\\
&|0,-\sqrt{2\alpha}\rangle_{7,8}|0,\sqrt{2\alpha},\rangle_{9,10}|\sqrt{2\alpha},0\rangle_{11,12}|-\alpha,\alpha,-\alpha\rangle_{4,5,6}+|0,-\sqrt{2\alpha}\rangle_{7,8}|0,\sqrt{2\alpha}\rangle_{9,10}|0,\sqrt{2\alpha}\rangle_{11,12}|-\alpha,\alpha,\alpha\rangle_{4,5,6}+\\
&|-\sqrt{2\alpha},0\rangle_{7,8}|\sqrt{2\alpha},0\rangle_{9,10}|\sqrt{2\alpha},0\rangle_{11,12}|\alpha,-\alpha,-\alpha\rangle_{4,5,6}+|-\sqrt{2\alpha},0\rangle_{7,8}|\sqrt{2\alpha},0\rangle_{9,10}|0,\sqrt{2\alpha}\rangle_{11,12}|\alpha,-\alpha,\alpha\rangle_{4,5,6}+\\
&|-\sqrt{2\alpha},0\rangle_{7,8}|0,\sqrt{2\alpha}\rangle_{9,10}|\sqrt{2\alpha},0\rangle_{11,12}|\alpha,\alpha,-\alpha\rangle_{4,5,6}+|-\sqrt{2\alpha},0\rangle_{7,8}|0,\sqrt{2\alpha}\rangle_{9,10}|0,\sqrt{2\alpha}\rangle_{11,12}|\alpha,\alpha,\alpha\rangle_{4,5,6})+\\
&A_2B_1(|0,-\sqrt{2\alpha},\rangle_{7,8}|\sqrt{2\alpha},0\rangle_{9,10}|0,-\sqrt{2\alpha}\rangle_{11,12}|-\alpha,-\alpha,-\alpha\rangle_{4,5,6}+|0,-\sqrt{2\alpha}\rangle_{7,8}|\sqrt{2\alpha},0\rangle_{9,10}|-\sqrt{2\alpha},0\rangle_{11,12}|-\alpha,-\alpha,\alpha\rangle_{4,5,6}+\\
&|0,-\sqrt{2\alpha}\rangle_{7,8}|0,\sqrt{2\alpha},\rangle_{9,10}|0,-\sqrt{2\alpha}\rangle_{11,12}|-\alpha,\alpha,-\alpha\rangle_{4,5,6}+|0,-\sqrt{2\alpha}\rangle_{7,8}|0,\sqrt{2\alpha}\rangle_{9,10}|-\sqrt{2\alpha},0\rangle_{11,12}|-\alpha,\alpha,\alpha\rangle_{4,5,6}+\\
&|-\sqrt{2\alpha},0\rangle_{7,8}|\sqrt{2\alpha},0\rangle_{9,10}|0,-\sqrt{2\alpha}\rangle_{11,12}|\alpha,-\alpha,-\alpha\rangle_{4,5,6}+|-\sqrt{2\alpha},0\rangle_{7,8}|\sqrt{2\alpha},0\rangle_{9,10}|-\sqrt{2\alpha},0\rangle_{11,12}|\alpha,-\alpha,\alpha\rangle_{4,5,6}+\\
&|-\sqrt{2\alpha},0\rangle_{7,8}|0,\sqrt{2\alpha}\rangle_{9,10}|0,-\sqrt{2\alpha}\rangle_{11,12}|\alpha,\alpha,-\alpha\rangle_{4,5,6}+|-\sqrt{2\alpha},0\rangle_{7,8}|0,\sqrt{2\alpha}\rangle_{9,10}|-\sqrt{2\alpha},0\rangle_{11,12}|\alpha,\alpha,\alpha\rangle_{4,5,6})+\\
&A_3B_0(|0,-\sqrt{2\alpha},\rangle_{7,8}|0,-\sqrt{2\alpha},\rangle_{9,10}|\sqrt{2\alpha},0\rangle_{11,12}|-\alpha,-\alpha,-\alpha\rangle_{4,5,6}+|0,-\sqrt{2\alpha}\rangle_{7,8}|0,-\sqrt{2\alpha}\rangle_{9,10}|0,\sqrt{2\alpha}\rangle_{11,12}|-\alpha,-\alpha,\alpha\rangle_{4,5,6}+\\
&|0,-\sqrt{2\alpha}\rangle_{7,8}|-\sqrt{2\alpha},0\rangle_{9,10}|\sqrt{2\alpha},0\rangle_{11,12}|-\alpha,\alpha,-\alpha\rangle_{4,5,6}+|0,-\sqrt{2\alpha}\rangle_{7,8}|-\sqrt{2\alpha},\rangle_{9,10}|0,\sqrt{2\alpha}\rangle_{11,12}|-\alpha,\alpha,\alpha\rangle_{4,5,6}+\\
&|-\sqrt{2\alpha},0\rangle_{7,8}|0,-\sqrt{2\alpha},0\rangle_{9,10}|\sqrt{2\alpha},0\rangle_{11,12}|\alpha,-\alpha,-\alpha\rangle_{4,5,6}+|-\sqrt{2\alpha},0\rangle_{7,8}|0,-\sqrt{2\alpha}\rangle_{9,10}|0,\sqrt{2\alpha},\rangle_{11,12}|\alpha,-\alpha,\alpha\rangle_{4,5,6}+\\
&|-\sqrt{2\alpha},0\rangle_{7,8}|-\sqrt{2\alpha},0\rangle_{9,10}|\sqrt{2\alpha},0\rangle_{11,12}|\alpha,\alpha,-\alpha\rangle_{4,5,6}+|-\sqrt{2\alpha},0\rangle_{7,8}|-\sqrt{2\alpha},0\rangle_{9,10}|0,\sqrt{2\alpha},\rangle_{11,12}|\alpha,\alpha,\alpha\rangle_{4,5,6})+\\
&A_3B_1(|0,-\sqrt{2\alpha},\rangle_{7,8}|0,-\sqrt{2\alpha},\rangle_{9,10}|0,-\sqrt{2\alpha},\rangle_{11,12}|-\alpha,-\alpha,-\alpha\rangle_{4,5,6}+|0,-\sqrt{2\alpha}\rangle_{7,8}|0,-\sqrt{2\alpha}\rangle_{9,10}|-\sqrt{2\alpha},0\rangle_{11,12}|-\alpha,-\alpha,\alpha\rangle_{4,5,6}+\\
&|0,-\sqrt{2\alpha}\rangle_{7,8}|-\sqrt{2\alpha},0\rangle_{9,10}|0,-\sqrt{2\alpha},\rangle_{11,12}|-\alpha,\alpha,-\alpha\rangle_{4,5,6}+|0,-\sqrt{2\alpha}\rangle_{7,8}|-\sqrt{2\alpha},\rangle_{9,10}|-\sqrt{2\alpha},0\rangle_{11,12}|-\alpha,\alpha,\alpha\rangle_{4,5,6}+\\
&|-\sqrt{2\alpha},0\rangle_{7,8}|0,-\sqrt{2\alpha},0\rangle_{9,10}|0,-\sqrt{2\alpha},\rangle_{11,12}|\alpha,-\alpha,-\alpha\rangle_{4,5,6}+|-\sqrt{2\alpha},0\rangle_{7,8}|0,-\sqrt{2\alpha}\rangle_{9,10}|-\sqrt{2\alpha},0\rangle_{11,12}|\alpha,-\alpha,\alpha\rangle_{4,5,6}+\\
&|-\sqrt{2\alpha},0\rangle_{7,8}|-\sqrt{2\alpha},0\rangle_{9,10}|0,-\sqrt{2\alpha},\rangle_{11,12}|\alpha,\alpha,-\alpha\rangle_{4,5,6}+|-\sqrt{2\alpha},0\rangle_{7,8}|-\sqrt{2\alpha},0\rangle_{9,10}|-\sqrt{2\alpha},0\rangle_{11,12}|\alpha,\alpha,\alpha\rangle_{4,5,6})]
\end{aligned}
\tag{A1}
$$

## Appendix B

**Table1.** All exhaustive possibilities for detection-events in Eq.(7a) along with requisite unitary operations for faithful ABQT and Partial ABQT Protocol.

| Photons detected in Alice lab. | | Photons detected in Bob lab | Heralded state-vectors along with their conditions $n_7, n_9, n_{11} = 0$ and $n_8, n_{10}, n_{12} \neq 0$(see Eq. (7a)) | Unitary operation in labs | | Faithful(F)/ Near Faithful (NF) |
|---|---|---|---|---|---|---|
| $n_7/\ n_8$ | $n_9/\ n_{10}$ | $n_{11}/\ n_{12}$ | | Alice | Bob | |
| Odd | Odd | Odd | $N_{A\dot{A}}(A_0\|\alpha,\alpha\rangle - A_1\|\alpha,-\alpha\rangle - A_2\|-\alpha,\alpha\rangle + A_3\|-\alpha,-\alpha\rangle)_{4,5} \otimes N_B(B_0\|\alpha\rangle - B_1\|-\alpha\rangle)_6$ | $D_6$ | $D_5 \otimes D_4$ | NF Eq.(12,13) |
| Odd | Odd | Even | $N_{A\dot{A}}(A_0\|\alpha,\alpha\rangle - A_1\|\alpha,-\alpha\rangle - A_2\|-\alpha,\alpha\rangle + A_3\|-\alpha,-\alpha\rangle)_{4,5} \otimes N_B(B_0\|\alpha\rangle + B_1\|-\alpha\rangle)_6$ | $I_6$ | $D_5 \otimes D_4$ | F/NF Eq.(14,15) |
| Odd | Even | Odd | $N_{A\dot{A}}(A_0\|\alpha,\alpha\rangle + A_1\|\alpha,-\alpha\rangle - A_2\|-\alpha,\alpha\rangle - A_3\|-\alpha,-\alpha\rangle)_{4,5} \otimes N_B(B_0\|\alpha\rangle - B_1\|-\alpha\rangle)_6$ | $D_6$ | $D_4$ | NF Eq.(16,17) |
| Even | Odd | Odd | $N_{A\dot{A}}(A_0\|\alpha,\alpha\rangle - A_1\|\alpha,-\alpha\rangle + A_2\|-\alpha,\alpha\rangle - A_3\|-\alpha,-\alpha\rangle)_{4,5} \otimes N_B(B_0\|\alpha\rangle - B_1\|-\alpha\rangle)_6$ | $D_6$ | $D_5$ | NF Eq(18,19) |
| Even | Even | Even | $N_{A\dot{A}}(A_0\|\alpha,\alpha\rangle + A_1\|\alpha,-\alpha\rangle + A_2\|-\alpha,\alpha\rangle + A_3\|-\alpha,-\alpha\rangle)_{4,5} \otimes N_B(B_0\|\alpha\rangle + B_1\|-\alpha\rangle)_6$ | $I_6$ | $I_5 \otimes I_4$ | F Eq.(20,21) |
| Even | Even | Odd | $N_{A\dot{A}}(A_0\|\alpha,\alpha\rangle + A_1\|\alpha,-\alpha\rangle + A_2\|-\alpha,\alpha\rangle + A_3\|-\alpha,-\alpha\rangle)_{4,5} \otimes N_B(B_0\|\alpha\rangle - B_1\|-\alpha\rangle)_6$ | $D_6$ | $I_5 \otimes I_4$ | NF/F Eq.(22,23) |
| Even | Odd | Even | $N_{A\dot{A}}(A_0\|\alpha,\alpha\rangle - A_1\|\alpha,-\alpha\rangle + A_2\|-\alpha,\alpha\rangle - A_3\|-\alpha,-\alpha\rangle)_{4,5} \otimes N_B(B_0\|\alpha\rangle + B_1\|-\alpha\rangle)_6$ | $I_6$ | $D_5$ | F/NF Eq.(24,25) |
| Odd | Even | Even | $N_{A\dot{A}}(A_0\|\alpha,\alpha\rangle + A_1\|\alpha,-\alpha\rangle - A_2\|-\alpha,\alpha\rangle - A_3\|-\alpha,-\alpha\rangle)_{4,5} \otimes N_B(B_0\|\alpha\rangle + B_1\|-\alpha\rangle)_6$ | $I_6$ | $D_4$ | F/NF Eq.(26,27) |

**Table 2.** All exhaustive possibilities for detection-events in Eq.(7b) along with requisite unitary operator for faithful/near-faithful ABQT Protocol.

| Photons detected in Alice lab. | | Photons detected in Bob lab. | Heralded state-vectors along with their conditions $n_7, n_9, n_{11} \neq 0$ and $n_8, n_{10}, n_{12} = 0$(see Eq.(7b) | Unitary operation in labs | | Faithful(F)/Near Faithful (NF) |
|---|---|---|---|---|---|---|
| $n_7/\, n_8$ | $n_9/\, n_{10}$ | $n_{11}/\, n_{12}$ | | Alice | Bob | |
| Odd | Odd | Odd | $N_{A\dot{A}}(A_0\lvert -\alpha,-\alpha\rangle - A_1\lvert -\alpha,\alpha\rangle - A_2\lvert \alpha,-\alpha\rangle + A_3\lvert \alpha,\alpha\rangle)_{4,5} \otimes N_B(B_0\lvert -\alpha\rangle - B_1\lvert \alpha\rangle)_6$ | $D_6P_6$ | $D_5 \otimes D_4(P_5 \otimes P_4)$ | NF Eq.(12,13) |
| Odd | Odd | Even | $N_{A\dot{A}}(A_0\lvert -\alpha,-\alpha\rangle - A_1\lvert -\alpha,\alpha\rangle - A_2\lvert \alpha,-\alpha\rangle + A_3\lvert \alpha,\alpha\rangle)_{4,5} \otimes (B_0\lvert -\alpha\rangle + B_1\lvert \alpha\rangle)_6$ | $P_6$ | $D_5 \otimes D_4(P_5 \otimes P_4)$ | F/NF Eq.(14,15) |
| Odd | Even | Odd | $N_{A\dot{A}}(A_0\lvert -\alpha,-\alpha\rangle + A_1\lvert -\alpha,\alpha\rangle - A_2\lvert \alpha,-\alpha\rangle - A_3\lvert \alpha,\alpha\rangle)_{4,5} \otimes N_B(B_0\lvert -\alpha\rangle - B_1\lvert \alpha\rangle)_6$ | $D_6P_6$ | $D_4\ (P_5 \otimes P_4)$ | NF Eq.(16,17) |
| Even | Odd | Odd | $N_{A\dot{A}}(A_0\lvert -\alpha,-\alpha\rangle - A_1\lvert -\alpha,\alpha\rangle + A_2\lvert \alpha,-\alpha\rangle - A_3\lvert \alpha,\alpha\rangle)_{4,5} \otimes N_B(B_0\lvert -\alpha\rangle - B_1\lvert \alpha\rangle)_6$ | $D_6P_6$ | $D_5\ (P_5 \otimes P_4)$ | NF Eq(18,19) |
| Even | Even | Even | $N_{A\dot{A}}(A_0\lvert -\alpha,-\alpha\rangle + A_1\lvert -\alpha,\alpha\rangle + A_2\lvert \alpha,-\alpha\rangle + A_3\lvert \alpha,\alpha\rangle)_{4,5} \otimes N_B(B_0\lvert -\alpha\rangle + B_1\lvert \alpha\rangle)_6$ | $P_6$ | $P_5 \otimes P_4$ | F Eq.(20,21) |
| Even | Even | Odd | $N_{A\dot{A}}(A_0\lvert -\alpha,-\alpha\rangle + A_1\lvert -\alpha,\alpha\rangle + A_2\lvert \alpha,-\alpha\rangle + A_3\lvert \alpha,\alpha\rangle)_{4,5} \otimes N_B(B_0\lvert -\alpha\rangle - B_1\lvert \alpha\rangle)_6$ | $D_6P_6$ | $P_5 \otimes P_4$ | NF/F Eq.(22,23) |
| Even | Odd | Even | $N_{A\dot{A}}(A_0\lvert -\alpha,-\alpha\rangle - A_1\lvert -\alpha,\alpha\rangle + A_2\lvert \alpha,-\alpha\rangle - A_3\lvert \alpha,\alpha\rangle)_{4,5} \otimes N_B(B_0\lvert -\alpha\rangle + B_1\lvert \alpha\rangle)_6$ | $P_6$ | $D_5\ (P_5 \otimes P_4)$ | F/NF Eq.(24,25) |
| Odd | Even | Even | $N_{A\dot{A}}(A_0\lvert -\alpha,-\alpha\rangle + A_1\lvert -\alpha,\alpha\rangle - A_2\lvert \alpha,-\alpha\rangle - A_3\lvert \alpha,\alpha\rangle)_{4,5} \otimes N_B(B_0\lvert -\alpha\rangle + B_1\lvert \alpha\rangle)_6$ | $P_6$ | $D_4\ (P_5 \otimes P_4)$ | F/NF Eq.(26,27) |

**Table 3.** All exhaustive possibilities for detection-events in Eq.(7c) along with requisite unitary operator for faithful/near-faithful ABQT Protocol.

| Photons detected in Alice lab. | | Photons detected in Bob lab. | Heralded state-vectors along with their conditions $n_7, n_9, n_{12} \neq 0$ and $n_8, n_{10}, n_{11} = 0$(see Eq. (7c)) | Unitary operation | | Faithful(F) /Near Faithful (NF) |
|---|---|---|---|---|---|---|
| $n_7/\ n_8$ | $n_9/\ n_{10}$ | $n_{11}/\ n_{12}$ | | Alice | Bob | |
| Odd | Odd | Odd | $N_{A\dot{A}}(A_0\|-\alpha,-\alpha\rangle - A_1\|-\alpha,\alpha\rangle - A_2\|\alpha,-\alpha\rangle + A_3\|\alpha,\alpha\rangle)_{4,5} \otimes N_B(B_0\|\alpha\rangle - B_1\|-\alpha\rangle)_6$ | $D_6$ | $(D_5 \otimes D_4)(P_5 \otimes P_4)$ | NF Eq.(12,13) |
| Odd | Odd | Even | $N_{A\dot{A}}(A_0\|-\alpha,-\alpha\rangle - A_1\|-\alpha,\alpha\rangle - A_2\|\alpha,-\alpha\rangle + A_3\|\alpha,\alpha\rangle)_{4,5} \otimes N_B(B_0\|\alpha\rangle + B_1\|-\alpha\rangle)_6$ | $I_6$ | $(D_5 \otimes D_4)(P_5 \otimes P_4)$ | F/NF Eq.(14,15) |
| Odd | Even | Odd | $N_{A\dot{A}}(A_0\|-\alpha,-\alpha\rangle + A_1\|-\alpha,\alpha\rangle - A_2\|\alpha,-\alpha\rangle - A_3\|\alpha,\alpha\rangle)_{4,5} \otimes N_B(B_0\|\alpha\rangle - B_1\|-\alpha\rangle)_6$ | $D_6$ | $D_4(P_5 \otimes P_4)$ | NF Eq.(16,17) |
| Even | Odd | Odd | $N_{A\dot{A}}(A_0\|-\alpha,-\alpha\rangle - A_1\|-\alpha,\alpha\rangle + A_2\|\alpha,-\alpha\rangle - A_3\|\alpha,\alpha\rangle)_{4,5} \otimes N_B(B_0\|\alpha\rangle - B_1\|-\alpha\rangle)_6$ | $D_6$ | $D_5(P_5 \otimes P_4)$ | NF Eq(18,19) |
| Even | Even | Even | $N_{A\dot{A}}(A_0\|-\alpha,-\alpha\rangle + A_1\|-\alpha,\alpha\rangle + A_2\|\alpha,-\alpha\rangle + A_3\|\alpha,\alpha\rangle)_{4,5} \otimes N_B(B_0\|\alpha\rangle + B_1\|-\alpha\rangle)_6$ | $I_6$ | $P_5 \otimes P_4$ | F Eq.(20,21) |
| Even | Even | Odd | $N_{A\dot{A}}(A_0\|-\alpha,-\alpha\rangle + A_1\|-\alpha,\alpha\rangle + A_2\|\alpha,-\alpha\rangle + A_3\|\alpha,\alpha\rangle)_{4,5} \otimes (B_0\|\alpha\rangle - B_1\|-\alpha\rangle)_6$ | $D_6$ | $P_5 \otimes P_4$ | NF/F Eq.(22,23) |
| Even | Odd | Even | $N_{A\dot{A}}(A_0\|-\alpha,-\alpha\rangle - A_1\|-\alpha,\alpha\rangle + A_2\|\alpha,-\alpha\rangle - A_3\|\alpha,\alpha\rangle)_{4,5} \otimes N_B(B_0\|\alpha\rangle + B_1\|-\alpha\rangle)_6$ | $I_6$ | $D_5(P_5 \otimes P_4)$ | F/NF Eq.(24,25) |
| Odd | Even | Even | $N_{A\dot{A}}(A_0\|-\alpha,-\alpha\rangle + A_1\|-\alpha,\alpha\rangle - A_2\|\alpha,-\alpha\rangle - A_3\|\alpha,\alpha\rangle)_{4,5} \otimes N_B(B_0\|\alpha\rangle + B_1\|-\alpha\rangle)_6$ | $I_6$ | $D_4(P_5 \otimes P_4)$ | F/NF Eq.(26,27) |

**Table 4.** All exhaustive possibilities for detection-events in Eq.(7d) along with requisite unitary operator for faithful/near-faithful ABQT Protocol.

| Photons detected in Alice lab. | | Photons detected in Bob lab. | Heralded state-vectors along with their conditions $n_7, n_{10}, n_{11} \neq 0$ and $n_8, n_9, n_{12} = 0$(see Eq.(7d)) | Unitary operation | | Faithful(F) /Near Faithful (NF) |
|---|---|---|---|---|---|---|
| $n_7/\ n_8$ | $n_9/\ n_{10}$ | $n_{11}/\ n_{12}$ | | Alice | Bob | |
| Odd | Odd | Odd | $N_{A\dot{A}}(A_0\|-\alpha,\alpha\rangle - A_1\|-\alpha,-\alpha\rangle - A_2\|\alpha,\alpha\rangle + A_3\|\alpha,-\alpha\rangle)_{4,5} \otimes N_B(B_0\|-\alpha\rangle - B_1\|\alpha\rangle)_6$ | $D_6 \otimes P_6$ | $(D_5 \otimes D_4)P_4$ | NF Eq.(12,13) |
| Odd | Odd | Even | $N_{A\dot{A}}(A_0\|-\alpha,\alpha\rangle - A_1\|-\alpha,-\alpha\rangle - A_2\|\alpha,\alpha\rangle + A_3\|\alpha,-\alpha\rangle)_{4,5} \otimes N_B(B_0\|-\alpha\rangle + B_1\|\alpha\rangle)_6$ | $P_6$ | $(D_5 \otimes D_4)P_4$ | F/NF Eq.(14,15) |
| Odd | Even | Odd | $N_{A\dot{A}}(A_0\|-\alpha,\alpha\rangle + A_1\|-\alpha,-\alpha\rangle - A_2\|\alpha,\alpha\rangle - A_3\|\alpha,-\alpha\rangle)_{4,5} \otimes N_B(B_0\|-\alpha\rangle - B_1\|\alpha\rangle)_6$ | $D_6 \otimes P_6$ | $D_4 \otimes P_4$ | NF Eq.(16,17) |
| Even | Odd | Odd | $N_{A\dot{A}}(A_0\|-\alpha,\alpha\rangle - A_1\|-\alpha,-\alpha\rangle + A_2\|\alpha,\alpha\rangle - A_3\|\alpha,-\alpha\rangle)_{4,5} \otimes N_B(B_0\|-\alpha\rangle - B_1\|\alpha\rangle)_6$ | $D_6 \otimes P_6$ | $D_5 \otimes P_4$ | NF Eq(18,19) |
| Even | Even | Even | $N_{A\dot{A}}(A_0\|-\alpha,\alpha\rangle + A_1\|-\alpha,-\alpha\rangle + A_2\|\alpha,\alpha\rangle + A_3\|\alpha,-\alpha\rangle)_{4,5} \otimes N_B(B_0\|-\alpha\rangle + B_1\|\alpha\rangle)_6$ | $P_6$ | $P_4$ | F Eq.(20,21) |
| Even | Even | Odd | $N_{A\dot{A}}(A_0\|-\alpha,\alpha\rangle + A_1\|-\alpha,-\alpha\rangle + A_2\|\alpha,\alpha\rangle + A_3\|\alpha,-\alpha\rangle)_{4,5} \otimes (B_0\|-\alpha\rangle - B_1\|\alpha\rangle)_6$ | $D_6 \otimes P_6$ | $P_4$ | NF/F Eq.(22,23) |
| Even | Odd | Even | $N_{A\dot{A}}(A_0\|-\alpha,\alpha\rangle - A_1\|-\alpha,-\alpha\rangle + A_2\|\alpha,\alpha\rangle - A_3\|\alpha,-\alpha\rangle)_{4,5} \otimes N_B(B_0\|-\alpha\rangle + B_1\|\alpha\rangle)_6$ | $P_6$ | $D_5 \otimes P_4$ | F/NF Eq.(24,25) |
| Odd | Even | Even | $N_{A\dot{A}}(A_0\|-\alpha,\alpha\rangle + A_1\|-\alpha,-\alpha\rangle - A_2\|\alpha,\alpha\rangle - A_3\|\alpha,-\alpha\rangle)_{4,5} \otimes N_B(B_0\|-\alpha\rangle + B_1\|\alpha\rangle)_6$ | $P_6$ | $D_4 \otimes P_4$ | F/NF Eq.(26,27) |

**Table 5.** All exhaustive possibilities for detection-events in Eq.(7e) along with requisite unitary operator for faithful/near-faithful ABQT Protocol.

| Photons detected in Alice lab. | | Photons detected in Bob lab. | Heralded state-vectors along with their conditions $n_7, n_{10}, n_{12} \neq 0$ and $n_8, n_9, n_{11} = 0$ (see Eq. (7e)) | Unitary operation | | Faithful(F)/ Near Faithful (NF) |
|---|---|---|---|---|---|---|
| $n_7/\ n_8$ | $n_9/\ n_{10}$ | $n_{11}/\ n_{12}$ | | Alice | Bob | |
| Odd | Odd | Odd | $N_{A\acute{A}}(A_0\|-\alpha,\alpha\rangle - A_1\|-\alpha,-\alpha\rangle - A_2\|\alpha,\alpha\rangle + A_3\|\alpha,-\alpha\rangle)_{4,5} \otimes N_B(B_0\|\alpha\rangle - B_1\|-\alpha\rangle)_6$ | $D_6$ | $(D_5 \otimes D_4)P_4$ | NF Eq.(12,13) |
| Odd | Odd | Even | $N_{A\acute{A}}(A_0\|-\alpha,\alpha\rangle + A_1\|-\alpha,-\alpha\rangle - A_2\|\alpha,\alpha\rangle + A_3\|\alpha,-\alpha\rangle)_{4,5} \otimes N_B(B_0\|\alpha\rangle + B_1\|-\alpha\rangle)_6$ | $I_6$ | $(D_5 \otimes D_4)P_4$ | F/NF Eq.(14,15) |
| Odd | Even | Odd | $N_{A\acute{A}}(A_0\|-\alpha,\alpha\rangle + A_1\|-\alpha,-\alpha\rangle - A_2\|\alpha,\alpha\rangle - A_3\|\alpha,-\alpha\rangle)_{4,5} \otimes N_B(B_0\|\alpha\rangle - B_1\|-\alpha\rangle)_6$ | $D_6$ | $D_4 \otimes P_4$ | NF Eq.(16,17) |
| Even | Odd | Odd | $N_{A\acute{A}}(A_0\|-\alpha,\alpha\rangle - A_1\|-\alpha,-\alpha\rangle + A_2\|\alpha,\alpha\rangle - A_3\|-\alpha,-\alpha\rangle)_{4,5} \otimes N_B(B_0\|\alpha\rangle - B_1\|-\alpha\rangle)_6$ | $D_6$ | $D_5 \otimes P_4$ | NF Eq(18,19) |
| Even | Even | Even | $N_{A\acute{A}}(A_0\|-\alpha,\alpha\rangle + A_1\|-\alpha,-\alpha\rangle + A_2\|\alpha,\alpha\rangle + A_3\|\alpha,-\alpha\rangle)_{4,5} \otimes N_B(B_0\|\alpha\rangle + B_1\|-\alpha\rangle)_6$ | $I_6$ | $P_4$ | F Eq.(20,21) |
| Even | Even | Odd | $N_{A\acute{A}}(A_0\|-\alpha,\alpha\rangle + A_1\|-\alpha,-\alpha\rangle + A_2\|\alpha,\alpha\rangle + A_3\|\alpha,-\alpha\rangle)_{4,5} \otimes N_B(B_0\|\alpha\rangle - B_1\|-\alpha\rangle)_6$ | $D_6$ | $P_4$ | NF/F Eq.(22,23) |
| Even | Odd | Even | $N_{A\acute{A}}(A_0\|-\alpha,\alpha\rangle - A_1\|-\alpha,-\alpha\rangle + A_2\|\alpha,\alpha\rangle - A_3\|\alpha,-\alpha\rangle)_{4,5} \otimes N_B(B_0\|\alpha\rangle + B_1\|-\alpha\rangle)_6$ | $I_6$ | $D_5 \otimes P_4$ | F/NF Eq.(24,25) |
| Odd | Even | Even | $N_{A\acute{A}}(A_0\|-\alpha,\alpha\rangle + A_1\|-\alpha,-\alpha\rangle - A_2\|\alpha,\alpha\rangle - A_3\|\alpha,-\alpha\rangle)_{4,5} \otimes N_B(B_0\|\alpha\rangle + B_1\|-\alpha\rangle)_6$ | $I_6$ | $D_4 \otimes P_4$ | F/NF Eq.(26,27) |

**Table 6.** All exhaustive possibilities for detection-events in Eq.(7f) along with requisite unitary operator for faithful/near-faithful ABQT Protocol.

| Photons detected in Alice lab. | | Photons detected in Bob lab. | Heralded state-vectors along with their conditions $n_7, n_{10}, n_{11} \neq 0$ and $n_8, n_9, n_{12} = 0$ (see Eq. (7f)) | Unitary operation | | Faithful(F)/ Near Faithful (NF) |
|---|---|---|---|---|---|---|
| $n_7/\, n_8$ | $n_9/\, n_{10}$ | $n_{11}/\, n_{12}$ | | Alice | Bob | |
| Odd | Odd | Odd | $N_{A\dot{A}}(A_0\|\alpha,-\alpha\rangle - A_1\|\alpha,\alpha\rangle - A_2\|-\alpha,-\alpha\rangle + A_3\|-\alpha,\alpha\rangle)_{4,5} \otimes N_B(B_0\|-\alpha\rangle - B_1\|\alpha\rangle)_6$ | $D_6 \otimes P_6$ | $(D_5 \otimes D_4)P_5$ | NF Eq.(12,13) |
| Odd | Odd | Even | $N_{A\dot{A}}(A_0\|\alpha,-\alpha\rangle - A_1\|\alpha,\alpha\rangle - A_2\|-\alpha,-\alpha\rangle + A_3\|-\alpha,\alpha\rangle)_{4,5} \otimes N_B(B_0\|-\alpha\rangle + B_1\|\alpha\rangle)_6$ | $P_6$ | $(D_5 \otimes D_4)P_5$ | F/NF Eq.(14,15) |
| Odd | Even | Odd | $N_{A\dot{A}}(A_0\|\alpha,-\alpha\rangle + A_1\|\alpha,\alpha\rangle - A_2\|-\alpha,-\alpha\rangle - A_3\|-\alpha,\alpha\rangle)_{4,5} \otimes N_B(B_0\|-\alpha\rangle - B_1\|\alpha\rangle)_6$ | $D_6 \otimes P_6$ | $D_4 \otimes P_5$ | NF Eq.(16,17) |
| Even | Odd | Odd | $N_{A\dot{A}}(A_0\|\alpha,-\alpha\rangle - A_1\|\alpha,\alpha\rangle + A_2\|-\alpha,-\alpha\rangle - A_3\|-\alpha,\alpha\rangle)_{4,5} \otimes N_B(B_0\|-\alpha\rangle - B_1\|\alpha\rangle)_6$ | $D_6 \otimes P_6$ | $D_5 \otimes P_5$ | NF Eq(18,19) |
| Even | Even | Even | $N_{A\dot{A}}(A_0\|\alpha,-\alpha\rangle + A_1\|\alpha,\alpha\rangle + A_2\|-\alpha,-\alpha\rangle + A_3\|-\alpha,\alpha\rangle)_{4,5} \otimes N_B(B_0\|-\alpha\rangle + B_1\|\alpha\rangle)_6$ | $P_6$ | $P_5$ | F Eq.(20,21) |
| Even | Even | Odd | $N_{A\dot{A}}(A_0\|\alpha,-\alpha\rangle + A_1\|\alpha,\alpha\rangle + A_2\|-\alpha,-\alpha\rangle + A_3\|-\alpha,\alpha\rangle)_{4,5} \otimes N_B(B_0\|-\alpha\rangle - B_1\|\alpha\rangle)_6$ | $D_6 \otimes P_6$ | $P_5$ | NF/F Eq.(22,23) |
| Even | Odd | Even | $N_{A\dot{A}}(A_0\|\alpha,-\alpha\rangle - A_1\|\alpha,\alpha\rangle + A_2\|-\alpha,-\alpha\rangle - A_3\|-\alpha,\alpha\rangle)_{4,5} \otimes N_B(B_0\|-\alpha\rangle + B_1\|\alpha\rangle)_6$ | $P_6$ | $D_5 \otimes P_5$ | F/NF Eq.(24,25) |
| Odd | Even | Even | $N_{A\dot{A}}(A_0\|\alpha,-\alpha\rangle + A_1\|\alpha,\alpha\rangle - A_2\|-\alpha,-\alpha\rangle - A_3\|-\alpha,\alpha\rangle)_{4,5} \otimes N_B(B_0\|-\alpha\rangle + B_1\|\alpha\rangle)_6$ | $P_6$ | $D_4 \otimes P_5$ | F/NF Eq.(26,27) |

**Table 7.** All exhaustive possibilities for detection-events in Eq.(7g) along with requisite unitary operator for faithful/near-faithful ABQT Protocol.

| Photons detectedin Alice lab. | | Photons detectedin Bob lab. | Heralded state-vectors along with their conditions $n_7, n_{10}, n_{12} \neq 0$ and $n_8, n_9, n_{11}$(see Eq. (7g)) | Unitary operation | | Faithful(F)/ Near Faithful (NF) |
|---|---|---|---|---|---|---|
| $n_7$/ $n_8$ | $n_9$/ $n_{10}$ | $n_{11}$/ $n_{12}$ | | Alice | Bob | |
| Odd | Odd | Odd | $N_{A\dot{A}}(A_0\|\alpha,-\alpha\rangle - A_1\|\alpha,\alpha\rangle - A_2\|-\alpha,-\alpha\rangle + A_3\|-\alpha,\alpha\rangle)_{4,5} \otimes N_B(B_0\|\alpha\rangle - B_1\|-\alpha\rangle)_6$ | $D_6$ | $(D_5 \otimes D_4)P_5$ | NF Eq.(12,13) |
| Odd | Odd | Even | $N_{A\dot{A}}(A_0\|\alpha,-\alpha\rangle + A_1\|\alpha,\alpha\rangle - A_2\|-\alpha,-\alpha\rangle + A_3\|-\alpha,\alpha\rangle)_{4,5} \otimes N_B(B_0\|\alpha\rangle + B_1\|-\alpha\rangle)_6$ | $I_6$ | $(D_5 \otimes D_4)P_5$ | F/NF Eq.(14,15) |
| Odd | Even | Odd | $N_{A\dot{A}}(A_0\|\alpha,-\alpha\rangle + A_1\|\alpha,\alpha\rangle - A_2\|-\alpha,-\alpha\rangle - A_3\|-\alpha,\alpha\rangle)_{4,5} \otimes N_B(B_0\|\alpha\rangle - B_1\|-\alpha\rangle)_6$ | $D_6$ | $D_4 \otimes P_4$ | NF Eq.(16,17) |
| Even | Odd | Odd | $N_{A\dot{A}}(A_0\|\alpha,-\alpha\rangle - A_1\|\alpha,\alpha\rangle + A_2\|-\alpha,-\alpha\rangle - A_3\|-\alpha,\alpha\rangle)_{4,5} \otimes N_B(B_0\|\alpha\rangle - B_1\|-\alpha\rangle)_6$ | $D_6$ | $D_5 \otimes P_5$ | NF Eq(18,19) |
| Even | Even | Even | $N_{A\dot{A}}(A_0\|-\alpha,\alpha\rangle + A_1\|-\alpha,-\alpha\rangle + A_2\|\alpha,\alpha\rangle + A_3\|\alpha,-\alpha\rangle)_{4,5} \otimes N_B(B_0\|\alpha\rangle + B_1\|-\alpha\rangle)_6$ | $I_6$ | $P_5 \otimes P_4$ | F Eq.(20,21) |
| Even | Even | Odd | $N_{A\dot{A}}(A_0\|-\alpha,\alpha\rangle + A_1\|-\alpha,-\alpha\rangle + A_2\|\alpha,\alpha\rangle + A_3\|\alpha,-\alpha\rangle)_{4,5} \otimes N_B(B_0\|\alpha\rangle - B_1\|-\alpha\rangle)_6$ | $D_6$ | $P_4$ | NF/F Eq.(22,23) |
| Even | Odd | Even | $N_{A\dot{A}}(A_0\|-\alpha,\alpha\rangle - A_1\|-\alpha,-\alpha\rangle + A_2\|\alpha,\alpha\rangle - A_3\|\alpha,-\alpha\rangle)_{4,5} \otimes N_B(B_0\|\alpha\rangle + B_1\|-\alpha\rangle)_6$ | $I_6$ | $D_5 \otimes P_4$ | F/NF Eq.(24,25) |
| Odd | Even | Even | $N_{A\dot{A}}(A_0\|-\alpha,\alpha\rangle + A_1\|-\alpha,-\alpha\rangle - A_2\|\alpha,\alpha\rangle - A_3\|\alpha,-\alpha\rangle)_{4,5} \otimes N_B(B_0\|\alpha\rangle + B_1\|-\alpha\rangle)_6$ | $I_6$ | $D_4 \otimes P_4$ | F/NF Eq.(26,27) |

**Table 8.** All exhaustive possibilities for detection-events in Eq.(7h) along with requisite unitary operator for faithful/near-faithful ABQT Protocol.

| Photons detectedin Alice lab. | | Photons detectedin Boblab. | Heralded state-vectors along with their conditions $n_7, n_{10}, n_{11} \neq 0$ and$n_8, n_9, n_{12} = 0$(see Eq. (7h)) | Unitary operation | | Faithful(F)/ Near Faithful (NF) |
|---|---|---|---|---|---|---|
| $n_7/\, n_8$ | $n_9/\, n_{10}$ | $n_{11}/\, n_{12}$ | | Alice | Bob | |
| Odd | Odd | Odd | $N_{A\dot{A}}(A_0\|\alpha,\alpha\rangle - A_1\|\alpha,-\alpha\rangle - A_2\|-\alpha,\alpha\rangle + A_3\|-\alpha,-\alpha\rangle)_{4,5} \otimes N_B(B_0\|-\alpha\rangle - B_1\|\alpha\rangle)_6$ | $D_6P_6$ | $D_5 \otimes D_4$ | NF Eq.(12,13) |
| Odd | Odd | Even | $N_{A\dot{A}}(A_0\|\alpha,\alpha\rangle + A_1\|\alpha,-\alpha\rangle - A_2\|-\alpha,\alpha\rangle + A_3\|-\alpha,-\alpha\rangle)_{4,5} \otimes N_B(B_0\|-\alpha\rangle + B_1\|\alpha\rangle)_6$ | $P_6$ | $D_5 \otimes D_4$ | F/NF Eq.(14,15) |
| Odd | Even | Odd | $N_{A\dot{A}}(A_0\|\alpha,\alpha\rangle + A_1\|\alpha,-\alpha\rangle - A_2\|-\alpha,\alpha\rangle - A_3\|-\alpha,-\alpha\rangle)_{4,5} \otimes N_B(B_0\|-\alpha\rangle - B_1\|\alpha\rangle)_6$ | $D_6P_6$ | $D_4$ | NF Eq.(16,17) |
| Even | Odd | Odd | $N_{A\dot{A}}(A_0\|\alpha,\alpha\rangle - A_1\|\alpha,-\alpha\rangle + A_2\|-\alpha,\alpha\rangle - A_3\|-\alpha,-\alpha\rangle)_{4,5} \otimes N_B(B_0\|-\alpha\rangle - B_1\|\alpha\rangle)_6$ | $D_6P_6$ | $D_5$ | NF Eq(18,19) |
| Even | Even | Even | $N_{A\dot{A}}(A_0\|\alpha,\alpha\rangle + A_1\|\alpha,-\alpha\rangle + A_2\|-\alpha,\alpha\rangle + A_3\|-\alpha,-\alpha\rangle)_{4,5} \otimes N_B(B_0\|-\alpha\rangle + B_1\|\alpha\rangle)_6$ | $P_6$ | $I_5 \otimes I_4$ | F Eq.(20,21) |
| Even | Even | Odd | $N_{A\dot{A}}(A_0\|\alpha,\alpha\rangle + A_1\|\alpha,-\alpha\rangle + A_2\|-\alpha,\alpha\rangle + A_3\|-\alpha,-\alpha\rangle)_{4,5} \otimes N_B(B_0\|-\alpha\rangle - B_1\|\alpha\rangle)_6$ | $D_6P_6$ | $I_5 \otimes I_4$ | NF/F Eq.(22,23) |
| Even | Odd | Even | $N_{A\dot{A}}(A_0\|\alpha,\alpha\rangle - A_1\|\alpha,-\alpha\rangle + A_2\|-\alpha,\alpha\rangle - A_3\|-\alpha,-\alpha\rangle)_{4,5} \otimes N_B(B_0\|-\alpha\rangle + B_1\|\alpha\rangle)_6$ | $P_6$ | $D_5$ | F/NF Eq.(24,25) |
| Odd | Even | Even | $N_{A\dot{A}}(A_0\|\alpha,\alpha\rangle + A_1\|\alpha,-\alpha\rangle - A_2\|-\alpha,\alpha\rangle - A_3\|-\alpha,-\alpha\rangle)_{4,5} \otimes N_B(B_0\|-\alpha\rangle + B_1\|\alpha\rangle)_6$ | $P_6$ | $D_4$ | F/NF Eq.(26,27) |